\documentclass[a4paper,11pt]{article}
\pdfoutput=1 
\usepackage{jcappub} 

\usepackage[utf8]{inputenc}
\usepackage[T1]{fontenc}
\usepackage{epsfig}
\usepackage{version}
\usepackage{amsmath}
\usepackage[makeroom]{cancel}	
\usepackage{amssymb}
\usepackage{lineno}	
\usepackage{wrapfig}
\usepackage{float}	
\usepackage{pdfpages}
\usepackage{makeidx}
\usepackage{graphicx}
\usepackage{bigfoot}		

\makeindex

\newcommand{\beq} {\begin{equation}}
\newcommand{\beqn} {\begin{eqnarray}}
\newcommand{\eeq} {\end{equation}}
\newcommand{\hmpc} {h^{-1}\mathrm{Mpc}}
\newcommand{\eeqn} {\end{eqnarray}}

\newcommand{\mbf} {\boldsymbol}
\newcommand{\Hone}{{\rm HI}}
\newcommand{\Fbar}{\overline{F}}

\newcommand{\lrf}{\lambda_{\rm RF}}

\newcommand{\apar}{\alpha_{\parallel}}
\newcommand{\aperp}{\alpha_{\perp}}
\newcommand{\rpar}{r_{\parallel}}
\newcommand{\rperp}{r_{\perp}}
\newcommand{\blya}{b_{\rm Ly\alpha}}
\newcommand{\biasetalya}{b_{\rm \eta Ly\alpha}}
\newcommand{\betalya}{\beta_{\rm Ly\alpha}}

\newcommand{\lcdm}{$\Lambda$CDM}
\newcommand{\Poned}{P^{1d}}
\newcommand{\Ponedf}{P^{1d}_f}

\def\lya{Ly$\alpha$}
\def\lyma{{\rm Ly}\alpha}
\def\agp{a_{\rm GP}}
\def\bgp{b_{\rm GP}}
\def\cgp{c_{\rm GP}}
\def\zt{z_t} 

\title{Mock data sets for the Eboss and DESI Lyman-$\alpha$ forest surveys}
\author[a]{Thomas Etourneau,}
\author[a,1]{Jean-Marc~Le~Goff,\note{Corresponding author.}}
\author[a,2]{James~Rich,\note{Corresponding author.}}
\author[b]{Ting~Tan,}
\author[c,3]{Andrei~Cuceu,\note{NASA Einstein Fellow.}}
\author[d]{ S.~Ahlen,}
\author[a]{E.~Armengaud,}
\author[e]{ D.~Brooks,}
\author[f]{ T.~Claybaugh,}
\author[g]{ A.~de la Macorra,}
\author[e]{ P.~Doel,}
\author[h]{ A.~Font-Ribera,}
\author[i,j]{ J.~E.~Forero-Romero,}
\author[f]{ S.~Gontcho A Gontcho,}
\author[k,l]{ A.~X.~Gonzalez-Morales,}
\author[l]{ H.~K.~Herrera-Alcantar,}
\author[c,m]{ K.~Honscheid,}
\author[f]{ T.~Kisner,}
\author[f]{ M.~Landriau,}
\author[h,n]{ M.~Manera,}
\author[c,o]{ P.~Martini}
\author[h,p]{ R.~Miquel}
\author[g]{A.~Muñoz-Gutiérrez}
\author[q]{ J.~Nie}
\author[r]{ I.~P\'erez-R\`afols}
\author[f,s]{ C.~Poppett}
\author[a,t]{ C.~Ravoux}
\author[u]{ M.~Rezaie}
\author[v]{ G.~Rossi}
\author[w]{ E.~Sanchez}
\author[x]{ M.~Schubnell}
\author[b]{ J.~Stermer}
\author[x]{ G.~Tarl\'{e}}
\author[y,z]{ M.~Walther}
\author[q]{ Z.~Zhou}

\affiliation[a]{IRFU, CEA, Universit\'{e} Paris-Saclay, F-91191 Gif-sur-Yvette, France}
\affiliation[b]{Sorbonne Universit\'{e}, CNRS/IN2P3, Laboratoire de Physique Nucl\'{e}aire et de Hautes Energies (LPNHE), FR-75005 Paris, France}
\affiliation[c]{Center for Cosmology and AstroParticle Physics, The Ohio State University, 191 West Woodruff Avenue, Columbus, OH 43210, USA}
\affiliation[d]{ Physics Dept., Boston University, 590 Commonwealth Avenue, Boston, MA 02215, USA} 
\affiliation[e]{ Department of Physics \& Astronomy, University College London, Gower Street, London, WC1E 6BT, UK}        
\affiliation[f]{ Lawrence Berkeley National Laboratory, 1 Cyclotron Road, Berkeley, CA 94720, USA}           
\affiliation[g]{ Instituto de F\'{\i}sica, Universidad Nacional Aut\'{o}noma de M\'{e}xico, Cd. de M\'{e}xico C.P. 04510, M\'{e}xico} 
\affiliation[h]{ Institut de F\'{i}sica d’Altes Energies (IFAE), The Barcelona Institute of Science and Technology, Campus UAB, 08193 Bellaterra Barcelona, Spain}
\affiliation[i]{ Departamento de F\'isica, Universidad de los Andes, Cra. 1 No. 18A-10, Edificio Ip, CP 111711, Bogot\'a, Colombia} 
\affiliation[j]{  Observatorio Astron\'omico, Universidad de los Andes, Cra. 1 No. 18A-10, Edificio H, CP 111711 Bogot\'a, Colombia}
\affiliation[k]{ Consejo Nacional de Ciencia y Tecnolog\'{\i}a, Av. Insurgentes Sur 1582. Colonia Cr\'{e}dito Constructor, Del. Benito Ju\'{a}rez C.P. 03940, M\'{e}xico D.F. M\'{e}xico}
\affiliation[l]{ Departamento de F\'{i}sica, Universidad de Guanajuato - DCI, C.P. 37150, Leon, Guanajuato, M\'{e}xico}
\affiliation[m]{ Department of Physics, The Ohio State University, 191 West Woodruff Avenue, Columbus, OH 43210, USA}
\affiliation[n]{ Departament de F\'{i}sica, Serra H\'{u}nter, Universitat Aut\`{o}noma de Barcelona, 08193 Bellaterra (Barcelona), Spain}
\affiliation[o]{ Instituci\'{o} Catalana de Recerca i Estudis Avan\c{c}ats, Passeig de Llu\'{\i}s Companys, 23, 08010 Barcelona, Spain}      
\affiliation[p]{ Instituci\'{o} Catalana de Recerca i Estudis Avan\c{c}ats, Passeig de Llu\'{\i}s Companys, 23, 08010 Barcelona, Spain} 
\affiliation[q]{ National Astronomical Observatories, Chinese Academy of Sciences, A20 Datun Rd., Chaoyang District, Beijing, 100012, P.R. China}  
\affiliation[r]{ Departament de F\'isica, EEBE, Universitat Polit\`ecnica de Catalunya, c/Eduard Maristany 10, 08930 Barcelona, Spain}      
\affiliation[s]{ Space Sciences Laboratory, University of California, Berkeley, 7 Gauss Way, Berkeley, CA  94720, USA}
\affiliation[t]{ Aix Marseille Univ, CNRS/IN2P3, CPPM, Marseille, France}            
\affiliation[u]{ Department of Physics, Kansas State University, 116 Cardwell Hall, Manhattan, KS 66506, USA}      
\affiliation[v]{ Department of Physics and Astronomy, Sejong University, Seoul, 143-747, Korea}            
\affiliation[w]{ CIEMAT, Avenida Complutense 40, E-28040 Madrid, Spain} 
\affiliation[x]{ Department of Physics, University of Michigan, Ann Arbor, MI 48109, USA}                               
\affiliation[y]{ Excellence Cluster ORIGINS, Boltzmannstrasse 2, D-85748 Garching, Germany}
\affiliation[z]{ University Observatory, Faculty of Physics, Ludwig-Maximilians-Universit\"{a}t, Scheinerstr. 1, 81677 M\"{u}nchen, Germany}

\emailAdd{jmlegoff@cea.fr}
\emailAdd{james.rich@cea.fr}

\date{Received xx xx 2021 / accepted  xx xx 2021}

\abstract{We present a  publicly-available code to generate sets of mock Lyman-$\alpha$ (\lya) forest data 
that have realistic large-scale correlations including those due to the Baryonic Acoustic Oscillations (BAO). 
The primary purpose of these mocks is to test the analysis procedures of the Extended Baryon Oscillation Survey (eBOSS) and the Dark Energy Spectroscopy Instrument (DESI) surveys.
The transmitted flux fraction, $F(\lambda)$,
of background quasars due to \lya\ absorption in the intergalactic medium (IGM)
is simulated using
the Fluctuating Gunn-Petterson Approximation 
(FGPA) applied to
Gaussian random fields produced through 
the use of fast Fourier transforms (FFT). 
The output includes the IGM-\lya\ transmitted flux fraction along quasar lines of sight and a catalog of high-column-density systems appropriately placed at high-density regions of the IGM.
This output  serves as input to additional code that superimposes
the  IGM tranmission on 
realistic quasar spectra, adds absorption by
high-column-density systems  and metals, 
and simulates instrumental transmission and noise. 
Redshift space distortions (RSD) of the flux correlations are implemented by including the 
large-scale velocity-gradient field in the
FGPA
resulting in a correlation function of $F(\lambda)$ that can be accurately predicted.  
One hundred realizations have been produced over the 14,000 deg$^2$ 
DESI survey footprint with 100 quasars per deg$^{2}$.
The analysis of these realizations shows that the correlations of $F(\lambda)$ follows the prediction within the accuracy of eBOSS survey.
The most time-consuming part of the mock production occurs before application of the FGPA,
and the existing pre-FGPA forests can be used to
easily produce new mock sets with modified redshift-dependent bias parameters or observational conditions.}

\begin{document}
\toccontinuoustrue
\maketitle
\flushbottom


\section{Introduction}

The Lyman-$\alpha$ (\lya) forest is a series of absorption features visible
in quasar spectra due to the presence of neutral hydrogen (\Hone)
in the intergalactic medium (IGM).
The Baryon Oscillation Spectroscopy Survey (BOSS)
\cite{2013AJ....145...10D} 
and the extended Baryon Oscillation Spectroscopy Survey (eBOSS) \citep{2016AJ....151...44D}
obtained the spectra of over $\approx200,000$ forests.
This permitted the  study of cosmological
large-scale structure (LSS)  through correlations of absorption between
different forests
\citep{2011SlosarBOSS,2013BuscaBOSS,2013SlosarDR9,2015DelubacDR11,2017BautistaDR12,2019deSainteAgatheDR14}
and through
correlations between forests and quasars
\citep{Font+13,Font+14,dMdB+17,2019BlomqvistDR14}.
These so-called 3d correlations
have been used to study baryonic acoustic oscillations 
(BAO)
with the most recent results found by a combination
of BOSS and eBOSS data~\cite{dMdB+20}
yielding
constraints on the parameters of 
cosmological models~\cite{eBOSS21}.
The 3d correlations have also been used 
to study the clustering of Damped \lya~systems (DLAs)
\citep{Font+12dla,DLAbias+18} 
and of the circumgalactic medium \citep{SBLA+22}.
Finally, correlations
within individual forests  allow the measurement of 
the so-called 1d power spectrum.
yielding constraints on the amplitude and shape
of the power spectrum at high redshift  \citep{Croft+98,McDonald+2000,Chabanier:2018rga,Ravoux+2023,Kara+2023}.
All of these studies are being continued with the currently
running Dark Energy Spectroscopy Instrument \cite{DESI:2022xcl}
which is expected to make considerable improvements in precision
of cosmological parameters \cite{DESI+2023}.

Most studies of large-scale structure 
have used galaxies as discrete tracers of the cosmological
density field.
In contrast to galaxies,  \lya\ absorption is a continuous probe of the density 
and a single quasar spectra provides a proxy for matter density along the quasar line-of-sight. 
On the other hand due to the relation between the transmitted flux fraction and the \lya\ optical depth, $F= \exp ( -\tau)$, the bias is negative and its absolute value is much smaller than unity.

Mock data sets are needed to test the analysis procedures
that lead to the \lya\ forest correlaions.
These mock data sets must cover hundreds of $h^{-3}{\rm Gpc^3}$,
while the absorption in a quasar spectrum  probes the intergalactic medium at the Jean's scale, i.e.~around 100$h^{-1}$ kpc. It is not possible to perform hydrodynamical or N-body simulations with such volumes and resolutions. 
We  are therefore limited to generating Gaussian random fields with a correlation close to that of the data.
The state-of-the-art mock sets presented here are not yet sufficiently realistic to be explicitly used in the analysis of real data by, for instance, providing a usable covariance matrix as is done in galaxy data sets.
Rather, they are used to demonstrate that the analysis procedure leads to estimates of physical parameters that are near those of the model used to generate the mocks and that the reported uncertainties are consistent with the variations observed between independent mock sets.  Most importantly, we want to demonstrate that the BAO peak position is correctly determined on average with a precision at the 1\% level.  The other physical parameters related to the broadband shape of the correlation functions are of less importance and we will see that the mocks confirm the accuracy of their determination only at the $10\%$ level.
The mocks are also used to  study astrophysical effects, such as those due to the presence of high column density (HCD) systems or metal lines in the \lya\ forest. 

The  BOSS survey used two sets of mocks, 
the first of which \cite{Font+12,Bautista+15}
used a Cholesky decomposition of the correlation matrix between pairs for forest pixels 
and generated a 3d correlated field only along the quasar lines-of-sight. 
This approach allows for the generation of mock data sets with whatever desired correlation function, e.g.~including the significant non-linear effects on small scales.
However, it results in no cross-correlation between the  flux transmission 
in the \lya\ pixels and the quasar positions,
so these mocks were  not useful for studies of
the forest-quasar cross-correlation

In order to include quasar-forest correlations, 
the second set of BOSS mocks 
\citep{dMdB+17}
was produced for the final BOSS analysis. 
They were based on a simpler approach, originally developed to study BOSS \lya\ analysis feasibility \citep{LeGoff+11}.
The Gaussian-random-field defining density fluctuations
was generated on a 3-dimensional grid with spacing 3.2 $h^{-1}$Mpc, defining 
a parallelepipedic box with 
cubic ``voxels'' of volume $(3.2\,h^{-1}{\rm Mpc})^3$.
Redshift space distortion (RSD) were implemented using the
corresponding velocity grid.
Real forest pairs have fixed angular separation so the
transverse separation of pairs depends on wavelength.
In these mocks, 
the approximation of parallel lines-of-sight was made,
so the separation is wavelength independent.
The positions of the quasars were originally~\cite{LeGoff+11} drawn randomly over the box, so for cross-correlation studies \citep{dMdB+17} the code was modified such that the quasar positions are selected in voxels with a density above a given threshold.

The eBOSS auto-correlation~\citep{dSA+19} analysis 
using SDSS data release 14 (DR14)
made use of BOSS Cholesky mocks~\cite{Font+12}
while the cross-correlation~\citep{Blomqvist+19} analysis used the second set of BOSS mocks with cross-correlation but the approximation of parallel lines-of-sight~\cite{dMdB+17}.  

Two mock projects were developed for the final eBOSS 
analysis using SDSS data release 16 (DR16) and for DESI. 
They use basically the same approach as the second set of BOSS mocks with the main improvement that the  lines-of-sight are no longer parallel.
The first project,  described in \citep{Farr+19}
produced what are referred to informally as "London mocks" or "Ly$\alpha$Colore mocks". 
This project used the same approach as the second set of BOSS mocks to select quasar locations and implement the RSD.
They were used to validate the analysis of the final eBOSS BAO study
\cite{dMdB+20}.
The second project presented here produced 
what are referred to informally as "Saclay mocks". 
In this project we have implemented RSD by including the 
large-scale velocity gradient fluctuations in the FGPA, allowing for a prediction of the mock correlation function. 
We use a lognormal field to select quasar locations
in a way that 
significantly improves the quasar auto-correlation at small separation.
These mocks have been used in the determination of the eBOSS DLA detection
efficiency \cite{DLAcatalogdr16}
and to study the correlation of voids and the \lya~forest \cite{Ravoux+2022}.

The production of realistic mocks is a four-step process.
The first step  is the calculation of IGM-\lya\ transmission field along lines of sight to quasars.
The second step is to add additional absorption due to HCDs and to metals.
The third is to apply this transmission on realistic
quasar spectra.
Finally, the fourth step introduces instrumental noise and atmospheric
transmission.
This paper is primarily concerned with the IGM transmission
of the first step.
The remaining steps are briefly described here but more completely in \cite{Herrera_2024}.

Section 2 explains how the mock data sets are produced, using a
Gaussian random field to generate the  IGM absorption field
and quasar positions.
The procedures for generating HCDs
and absorption by metals are also
briefly discussed.
Section 3 describes the implementation and the production of these mocks. Section 4 presents the results of the analysis of the mock sets
that contain no HCDs or metals.
The analysis of mock sets that include absorption by HCDs will be described
in a forthcoming publication [Tan et al, in preparation].
Conclusions are drawn in Sect. \ref{section:conclusions}. 
Several appendices describe some technical issues that were not detailed in section 2.

\begin{table*}
    \caption{ 
    Cosmological parameters \cite{Planck+2018}
    of the flat \lcdm~model used in the Mock generation.
    }
    \label{table:cosmoparams}
    \begin{tabular}{l | r | l }
    Parameter & value \\
    \hline
$\Omega_M$ & 0.3147  & density of non-relativistic matter \\
$\Omega_c h^2$ & 0.1198 & density of cold dark matter\\
$\Omega_b h^2$ & 0.02233 & density of baryons \\
$h$ & 0.6737 & Expansion rate 
$H_0/(100\,{\rm km^{-1}\,s^{-1}\,Mpc^{-1})}$\\
$A_s \times 10^{10}$ & 3.043 & Amplitude of scalar perterbations\\
$n_s$ & 0.9652 & spectral index of scalar perturbations \\
         \end{tabular}
\end{table*}

    \begin{figure}[H] 
\centering
\includegraphics[scale=0.6]{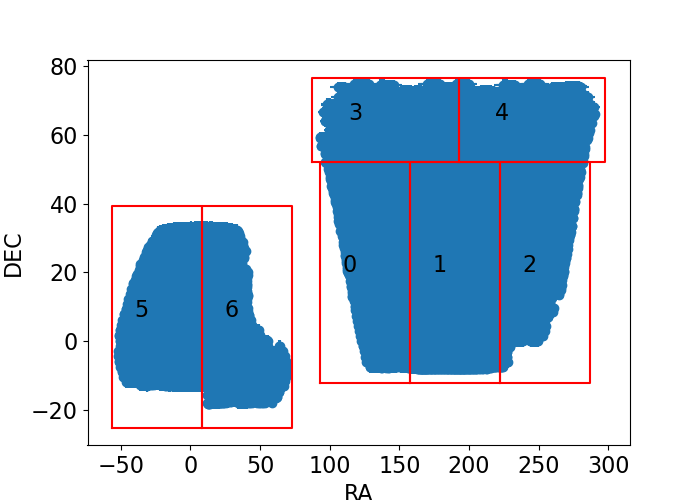} 
    \caption{
    The position in (ra,dec) of the
    seven "chunks" containing the density, velocity,
    and velocity gradient boxes used to generate the mocks.
    The blue region is the  DESI footprint (which
    includes the eBOSS footprint).  Each box
    contains $2560\times2560\times1536$ cubic voxels
    of volume $\rm (2.19 h^{-1} Mpc)^3$.
    The radial dimension has 
    1536 grid steps and  covers
    the redshift range $1.5<z<3.8$.
    The boxes are adjacent at $z=3.8$ and therefore
    overlap for $z<3.8$. 
    } 
    \label{fig:chunks}
    \end{figure}

\section{Mock generation}

We generate density and velocity fields in the DESI
footprint by combining "boxes" of volume 
$V\approx105({\rm Gpc}/h)^3$ containing
$2560\times2560\times1536$
cubic voxels of 2.19 $h^{-1}$ Mpc sides.
The positioning of the boxes is shown in Fig.~\ref{fig:chunks}.

For each box, an independent normal Gaussian random variable, $\delta(\mbf r)$, is generated in each voxel of the box, which results in a constant power spectrum of the box, $P(k)=V_{vox}$, the volume of a voxel. 
We perform a 3d Fourier transform using a fast 
Fourier transform algorithm (FFT) \cite{1965-cooley},
which results in a box in $k$-space with $\hat\delta(-\mbf k) =  [\hat\delta(\mbf k)]^*$, 
where we place hats on fields in $k$-space~\footnote{We could have directly generated the $\hat\delta(\mbf k)$ box in $k-$space saving one FFT. 
We chose to use the more straight-forward procedure of
generating first the field in real space since at any rate
we need to perform 13 inverse FFTs of the k-space field
to generate
the associated real-space fields (as explained below), so we would have saved only 7\% of the CPU time associated with the FFTs.}.
For a power spectrum $P_0(\mbf k)$, multiplying each mode $\hat\delta(\mbf k)$ by $\sqrt{P_0(\mbf k)/V_{vox}}$ and applying inverse FFT results in a field in real space with power spectrum $P_0$. 
For a given set of $\hat\delta(\mbf k)$ we thereby generate four physical fields with
differing but related power spectra:  the matter density  field at $z=0$ (Eq.~\ref{eq:deltaL})
and three lognormal quasar-density fields at different redshifts (section \ref{sec:qso}).
We also define, in $k$-space, 
boxes filled with the corresponding fields 
for the velocity (Eq.~\ref{eq:velocity}) 
and velocity gradient (Eq.~\ref{eq:eta}) 
and apply inverse FFT to obtain the corresponding fields in real space. 
The details of these fields are given the following subsections.

\subsection{Matter-density, velocity and velocity-gradient boxes}
\label{sec:matter}

For each mode $\mbf k$ we introduce 
\beq
\label{eq:deltaL}
\hat\delta_m (\mbf k)= \sqrt{\frac{P_m(\mbf k)}{V} }\; \hat\delta(\mbf k) \;,
\eeq
where $P_m$ is the matter power spectrum at $z=0$ obtained using 
CAMB \citep{CAMB+2000} 
with flat-\lcdm~parameters 
shown in Table~\ref{table:cosmoparams}.
This results in a correlated Gaussian random field, $\delta_m$, in real space with power spectrum $P_m$.

Within linear approximation, the velocity field in $\mbf k$-space reads
\beq
\hat v_j(\mbf k) = i f a H(z) \frac{k_j }{k^2} \; \hat \delta_m (\mbf k) \; ,
\label{eq:velocity}
\eeq 
 where $a=1/(1+z)$, $f = d \ln G / d \ln a$ is the linear growth rate of structure with $G$ the growth factor, and $H(z)$ is the Hubble parameter.
We produce three boxes that contain the three coordinates, $v_j$, of the velocity field in real space at $z=0$. 
After multiplying by a factor $faH/(faH)_{z=0}$ to go from $z=0$ to the considered quasar redshift, these boxes provide the radial peculiar velocity, $v_\parallel = u_j v^j$, for a quasar line-of-sight with unit vector $u_j$. This velocity modifies the cosmological redshift and generates the associated redshift-space distortions (RSD) of the quasar auto-correlation function.

Still within linear approximation the velocity-gradient 
fluctuation in $\mbf k$-space reads
 \beq
 \hat\eta_{pq} (\mbf k)= f\frac{k_p k_q}{k^2}\hat\delta_m(\mbf k) \; ,
 \label{eq:eta}
 \eeq 
By applying inverse Fourier transform we produce 6 boxes 
that contains the resulting velocity gradient components in real space. The line-of-sight velocity gradient is then $\eta_\parallel = u_p u_q \eta^{pq}$.

\subsection{Quasar generation}
\label{sec:qso}

Quasars are biased tracers of the matter distribution and
we wish to place quasars so that their power spectrum is close to the Kaiser power spectrum 
\beq
P_{QSO}(\mbf k,z) = b^2_{QSO}(z)(1+\beta_{QSO}(z) \mu_k^2)^2P_m(k,z),
\label{eq:PQSO}
\eeq
where $b_{QSO}$ and $\beta_{QSO}$ are the quasar bias and RSD parameters, $\mu_k=k_\parallel/k$, and $P_m$ is the matter power spectrum.
In Refs.~\citep{LeGoff+11} and~\citep{Farr+19}, 
candidates for quasar locations were uniformly drawn in cells where the matter field is above a given threshold, which is selected to produce the required bias. 
The selected peaks were then  Poisson sampled 
to follow the observed quasar redshift distribution.
Varying the threshold with redshift provides a $z$-dependent correlation function. 
As indicated by formula 5.102 in \cite{Padmanabhan93}, this procedure works well when the correlation function is $\xi\ll1$, i.e.~on large scales, but on small scales it results in a quasar auto-correlation function that is significantly larger than the linear prediction,
and also larger than the observed auto-correlation, as measured e.g. in Ref.~\cite{Laurent+17}.

To solve this problem, we used lognormal fields~\cite{ColesJones91}. 
We produce a quasar-density box, filled with a Gaussian field $\delta_q$
which, like $\delta_m$, is derived from the original $\hat\delta(\mbf k)$
but with mode amplidudes, $\hat\delta_q(\mbf k)$,
that are modified from the $\hat\delta_m(\mbf k)$
of eqn \ref{eq:deltaL}:
$\hat\delta_q (\mbf k)= \sqrt{P_q(\mbf k)/V}\; \hat\delta(\mbf k)$
where  $P_q(k)$ is the Fourier transform of $\ln[1+b_{QSO}^2(z_0) \xi_m(r,z_0)]$. 
We then draw a quasar in each voxel with a probability proportional to the lognormal field, $\exp(\delta_q)$. 
Once a voxel is selected to host a quasar, the quasar position is drawn uniformly within the voxel, as in Refs.~\citep{LeGoff+11} and~\citep{Farr+19}.
Ref. \cite{ColesJones91} showed that the correlation function of the lognormal field  $\exp(\delta_q)$
has the form 
$b_{QSO}^2(z_0) \xi_m(r,z_0)$.
The quasar velocity is taken  from Eq.~\ref{eq:velocity},
yielding a redshift parameter  $\beta_{QSO}=f/b_{QSO}$.

The quasar bias evolves with redshift as $b_{QSO}(z) = 3.7 \times((1+z)/(1+2.33))^{1.7}$ \citep{Laurent+17}.
The $z$-dependence of the quasar correlation function is implemented by producing three quasar-density boxes at $z_1=1.9$, $z_2=2.75$ and $z_3=3.6$ and interpolating between them, as described in appendix \ref{app:A}.


\subsection{Gaussian fields along lines of sight}

We next define a Gaussian field, $g$, defined 
along lines of sight to quasars and subdivided
into pixels of length $0.2\,\hmpc$.  The field has three components:
\beq 
 g =  \delta_L + \delta_S + \cgp(z) \eta_\parallel \; ,
\label{eq:gfield}
\eeq
The components $\delta_L$  and $\eta_\parallel$ describe the 
large-scale density and velocity-gradient fluctuations 
and are derived directly from the voxel values of $\delta_m$ and $\eta_{pq}$
given by equations \ref{eq:deltaL} and \ref{eq:eta}.
The parameter 
$c_{\mathrm GP}=\beta_{Ly\alpha} /f$ describes
the desired RSD due to the bias parameter $\beta_{Ly\alpha}$ and
growth rate $f\approx1$.
The component $\delta_S$ accounts for small-scale density fluctuations
not included in $\delta_L$.

If the values of $\delta_L$ and $\eta_\parallel$ at each pixel were defined by
the values of $\delta_m$ and $\eta_{pq}$ at the nearest voxel, there would be
discontinuities and then aliasing to small scales, i.e.~spurious power on $k$ larger than $k_{N}=\pi/a$ (Nyquist $k$), where $a$ is the side of the box voxels. 
To avoid this, the value of the spectra in a given pixel is defined as a Gaussian-kernel average over neighboring voxels. 
We use the voxel side as the standard-deviation of this Gaussian smoothing, which cuts the power spectrum at large $k$, making the aliasing terms at $k>k_N$ completely negligible. In practice we compute the Gaussian weighting including three voxels in each direction, i.e.~$7^3=343$ voxels, which means truncating the Gaussian at more than 3$\sigma$, which has negligible effect 
on the value of $\delta_L$ and $\eta_\parallel$.

The small effect of the Gaussian smoothing on the correlation function of $\delta_L$ can be computed and the even smaller effect of the averaging over voxel volumes can be approximated.
For the nominal voxel side used, $a=2.19 \, h^{-1}$ Mpc, the difference between the correlation functions of $\delta_L$ and matter on scales larger than about $10h^{-1}$ Mpc
is much less than the observational uncertainties for eBOSS final data (DR16).

While the $\delta_L$ pixels reflect the matter 3d correlation 
on large scales, 
the Mpc size of the voxels means that $\delta_L$ does not
contain
a significant amount of small scale fluctuations,
as detailed in appendix \ref{app:B}.
This results in a pixel variance and a 1d power spectrum of the field that are significantly lower than in real data~\footnote{The apparent contradiction between proper 3d correlation function and significantly smaller 1d power spectrum is discussed in the appendix.}.
Solving this issue by reducing the voxel size to the Jean's scale would require 3d FFT with more than $10^{13}$ voxels, which is not feasible.

The idea is then to add a small scale spectrum to the large scale spectrum, $\delta = \delta_L +\delta_S$, where $\delta_S$ is a random spectrum that is generated independently for each spectrum. It is therefore uncorrelated between different lines-of-sight and the 3d correlation function of $\delta$ is then that of $\delta_L$, keeping a 3d correlation function close to that of matter.
We use the same procedure to generate the $\delta_S$ spectrum in 1d as was used to generate the $\delta_m$ box in 3d:
a Gaussian field  is defined over the pixel range, and the Fourier components of this field are multiplied by the square root of the desired 1d power spectrum.  An inverse Fourier transform then defines $\delta_S$.
This allows us to give to $\delta_s$ the required 1d correlation on all scales to get a 1d power spectrum for the flux close to that of the data.

\subsection{FGPA}

\begin{figure}[H] 
\centering
\includegraphics[scale=0.6]{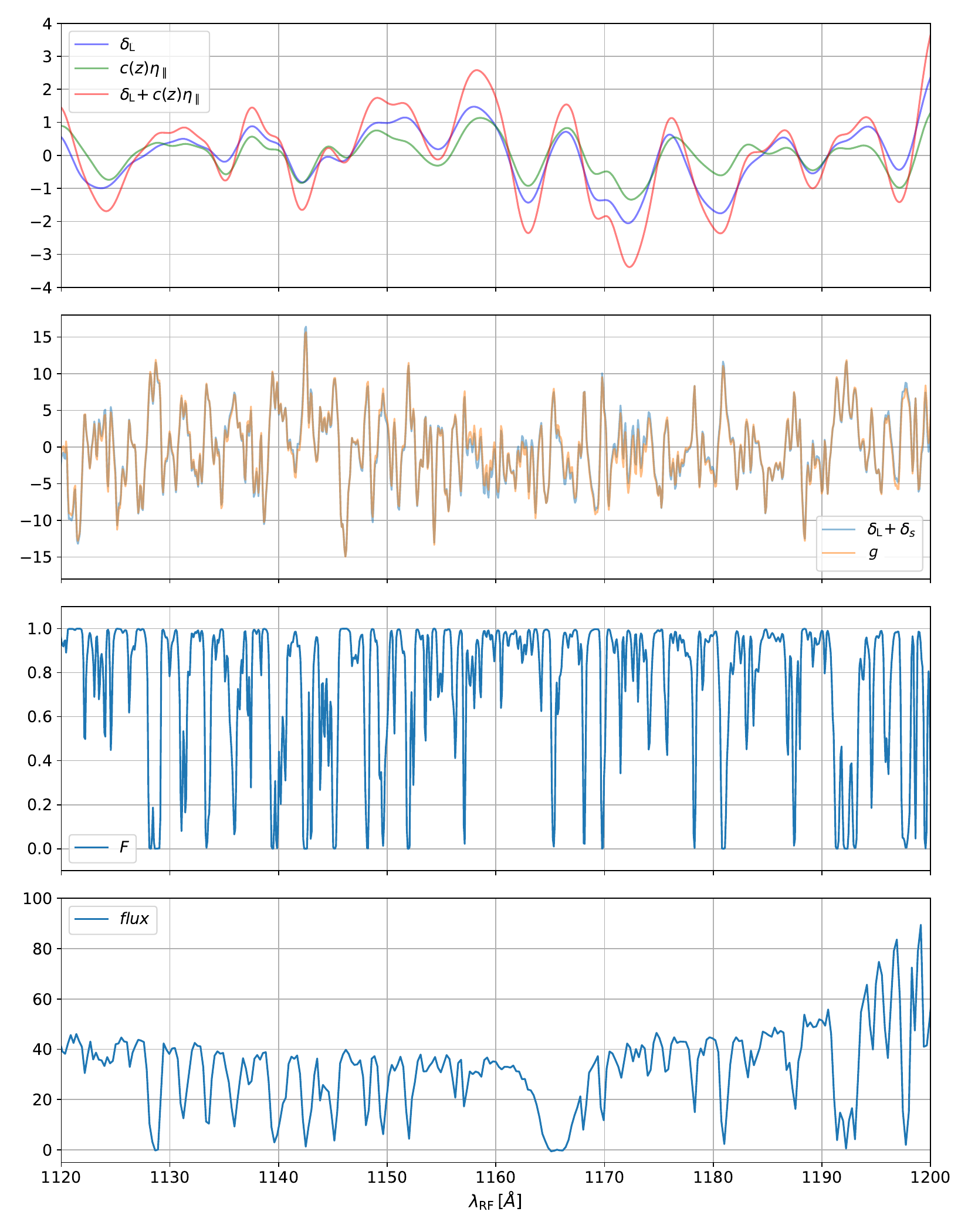}
\caption{
Steps in the production of mock forests showing 
for one representative forest the
fields as a function of quasar restframe wavelength, $\lrf$.
From top to bottom, the first graph shows the large-scale fields $\delta_L$, $c(z)\eta_\parallel$ and the sum of the two. The second presents the sum $\delta_L+\delta_S$ and the field $g=\delta_L+\delta_S +c(z)\eta_\parallel$. The third presents the transmitted flux fraction $F$ , and the last graph presents the spectrum
(the product of $F$ and the unabsorbed quasar spectrum)
A DLA has been added at $\lrf \approx 1165$A. 
}
\label{fig:spectrum}
\end{figure}

The basic idea behind the FGPA is that the optical depth for \lya\ absorption by the IGM is
 $\tau(z)\propto n_{\Hone}(z)/v^\prime(z)$,  where  $n_{\Hone}$ and $v^\prime$ are the neutral hydrogen density and velocity
 gradient at the  redshift where photons of a given observed wavelength can induce
  the \lya\ transition~\citep{Weinberg+99}.
This formula  is valid if when thermal broadening and turbulence can be neglected as indicated by hydrodynamical simulations of the IGM.
In addition, the simple physics that governs the ionization of the low density intergalactic medium implies that $ n_{\Hone} \propto   (1+\delta)^b$~\citep{HuiGnedin97}.
If one neglects peculiar velocity fluctuations, the lognormal approximation~\cite{ColesJones91, BiDavidsen97} then leads to $\tau \propto \exp (b \delta)$, and the more readily measured transmitted flux fraction is $F=\exp(-\tau)=\exp[ -a \exp (b\delta)]$ 
\footnote{this is also sometimes referred to as the FGPA, e.g.~Eq 2.5 in Ref.~\cite{Farr+19}.}

Including line-of-sight velocity gradient fluctuations, $\eta_\parallel$, within the lognormal approximation defines $F$ in terms of the Gaussian field $g(\lambda)$ (eqn. \ref{eq:gfield}):
 \beq 
     F(\lambda) = \exp [ -\agp(z) \exp( \bgp(z) g) ]  \quad  {\rm with} \quad  g(\lambda) =  \delta_L(\lambda) + \delta_S(\lambda) + \cgp(z) \eta_\parallel(\lambda) \; ,
\label{eq:FGPA}
\eeq
The additional bias redshift-dependent parameter, $\cgp$, would be a simple function of the
mean flux \cite{Seljak12,CieplakSlosar2016} were  it not for residual thermal effects.
We treat it, along with $\agp$, $\bgp$
and the power spectrum of $\delta_S$, 
as parameters that are chosen to reproduce
the observed statistical properties of $F$, 
as described in Sect.~\ref{sec:adjustparameters}.
For the wavelength $\lambda$ associated with the line-of-sight pixels,
we use the unperturbed cosmological redshift, i.e.~we neglect
the peculiar velocity in the wavelength calculation.
Mathematically, this ensures that the power spectrum of the $g$ field corresponds to the Kaiser formula (see section \ref{sec:prediction}).
Including the peculiar velocity in the wavelength calculation would result in a slightly different power spectrum.

Fig.~\ref{fig:spectrum} gives an example of a spectrum, with the different steps from $\delta_L$ to $F$.

We emphsize that while there is some physical justification of eqn.~(\ref{eq:FGPA}),
it should be just considered as a convenient 
way of generating mocks with the appropriate
correlations.

 \subsection{The expected auto-correlation function of the mocks}
\label{sec:prediction}

Our implementation of RSD differs from  Refs.~\citep{LeGoff+11} and~\citep{Farr+19},  which used the  FGPA without velocity gradient fluctuations and then 
introduced the RSD by displacing the \Hone\ according to the peculiar velocity.
Our introduction of the velocity gradient directly in the FGPA  has the nice feature that it allows us to predict the correlation function of the mocks. In this section we discuss this prediction, given the mock parameters and in particular the FGPA transformation.

Since the field $\delta_S$ is generated independently for each spectrum, the 3d correlation of $\delta_S$ vanishes, and including  $\delta_S$ in the field $g=\delta_L+\delta_S + \cgp(z) \eta_\parallel$ does not change its 3d correlation, it only adds noise.  
Eq.~\ref{eq:eta} implies that 
$\hat\eta_\parallel=f[(\mbf u . \mbf k)^2/k^2] \hat\delta = f \mu_k^2 \hat\delta$ and the power spectrum of the field $g$ is,
ignoring the noise from $\delta_S$
\beq
P_g ({\mbf k})= \left(1+f\cgp\mu_k^2\right)^2 \times P_m(k,z=0) W^2(\mbf k)\; ,
\label{eq:Pg}
\eeq
where the window $W(\mbf k)$ is the product of a Gaussian smearing window 
(with standard deviation $1/a$)
and the three windows that take into account the averaging of the field over the cubic voxels in the three directions, e.g.\ $W_x=\mathrm{sinc}(k_x a/2)$, where $a=2.19 h^{-1}$ Mpc is the voxel side. 
Replacing the product $W_x W_y W_z$ by the window of a sphere of diameter $a$ (that is entirely inside the voxel)~\footnote{The window is $3(\sin x-x\cos x)/x^3$ with $x=ka/2$} results in an overestimation of the product, while a sphere of diameter $\sqrt{3}a$ (that includes the whole voxel) results in an underestimation. We use the approximation $W_x W_y W_z \approx \mathrm{sinc}^3 (ka/(2\sqrt{3}))$, where we assume that in average $k_x=k_y=k_z=k/\sqrt{3}$, which nicely sits between the under- and over-estimation. This approximation depends only on $k$ and can be included in the definition of the power spectrum.  Eq.~\ref{eq:Pg} then corresponds to a Kaiser power spectrum with RSD parameter $\beta=f\cgp$ and the correlation function of $g$ can be obtained using formulas (5) to (9) of Ref.~\cite{Hamilton91}. 
It appears that the dominant window effect is the Gaussian one and the choice of the approximation for $W_x W_y W_z$ does not really matter.
(Fig.~\ref{fig:P1D} of Appendix~\ref{app:B}
illustrates this for the case of $\Poned$.)

The correlation function of $F$ is then deduced from that of $g$ using Eqn.~(2.6) of ref.~\cite{Font+12}. 
In Sect. \ref{sec::analraw} (Fig.~\ref{fig:xi_lya}),
we will see that
the prediction is correct to a good approximation, even if there is a small discrepancy along the line-of-sight, which we ascribe to our approximation for the product $W_x W_y W_z$.
In the range where we usually fit the data, $r>20 h^{-1}$ Mpc, the prediction agrees with the measurement within the error of one realization of eBOSS.

\subsection{Adjusting mock parameters}
\label{sec:adjustparameters}

To produce mocks with the desired properties, parameter tuning is performed for 5 values of the redshift, $\zt=1.8$, 2.2, 2.6, 3.0 and 3.6, and then interpolated for other values of $z$.
At a given $\zt$, the mock parameters include $\agp(\zt)$, $\bgp(\zt)$ and $\cgp(\zt)$ that appear in Eq.~\ref{eq:FGPA} 
and the power spectrum, $P_s(k,\zt)$, of the field $\delta_S$ over the range $[0,2]$ $h$Mpc$^{-1}$. 
The values of these parameters are chosen to produce the desired values 
of the mean transmitted flux fraction $\Fbar(z)$,
a 3d power spectrum that is a Kaiser power spectrum with bias parameters, $b_{\lyma}(z)$ 
and $\beta_{\lyma}(z)$, and a 1d flux power spectrum, $\Ponedf$, as measured by BOSS~\cite{Palanque+13}. 
We will see that the tuning procedure succeeds in
producing physically reasonable mocks while not yielding exactly the target parameters and correlations.

In practice, instead of  targeting a value of $b_{\lyma}$ 
for each $z_t$ we target an  effective bias, 
$b_{{\rm eff},\lyma}= b_{\lyma}(1+2\beta_{\lyma}/3+\beta_{\lyma}^2/5)^{1/2}$, 
which is proportional to the monopole of the correlation function~\cite{Hamilton91}.
To choose the target values we analyzed the eBOSS DR16 data 
following the same procedure as Ref.~\cite{dMdB+20}.
We found that the bias parameters are well fit by power laws
$b_{{\rm eff},\lyma}(z)=0.183[(1+z)/3.3]^{3.47}$ 
and $\beta_{\lyma}(z)=1.76[(1+z)/3.3]^{-2.32}$
(Fig.~\ref{fig:b_lya} of Sect. \ref{sec::analraw}).
We take these power laws as our targets.

For the mean transmitted flux fraction we use as target  
$\Fbar(z) = \exp(-0.0028(1+z)^{3.45})$, which is the parameterization of the measured $\Fbar(z)$ obtained by \cite{Calura+12}.

As already noted the RSD parameter of the $g$ field  is simply 
${f\cgp(\zt)}=\beta_{\lyma}(\zt)$. 
Given that $g$ in eq.~\ref{eq:FGPA} is a Gaussian field
(of variance $\sigma_g^2$), 
the values of $\Fbar$ and 
its variance $\sigma_F^2$ 
only depend on $\agp$ and $\sigma(\bgp  g)=\bgp  \sigma_g$~\cite{Font+12}. 
Therefore, we 
fit\footnote{We use the Minuit package~\cite{minuit}} $\agp$ and $\bgp\sigma_g$
to reach the target $\Fbar(\zt)$ and $\sigma_F(\zt)$, where the target for $\sigma_F(\zt)$ is obtained from the integral of our target $\Ponedf(k,\zt)$.

To fix $P_s(k)$, 
we use an iterative procedure. As an initial $P_s(k)$ we take the difference between the CAMB matter 1d power spectrum  
and the 1d power spectrum of the $\delta_L+\cgp \eta_\parallel$ field. 
(While this is a crude approximation since the CAMB
spectrum ignores small-scale effects, it is sufficient 
as an initial spectrum.)
The integral of this initial $P_s(k)$ 
gives  $\sigma_s$ (the dispersion of the field $\delta_s$) .
We can measure $\sigma_L$, $ \sigma_\eta$ and cov$(\delta_L,\eta_\parallel)$ directly on the produced boxes.
We, therefore, have all elements to compute
$\sigma_g^2 = \sigma_l^2 + \sigma_s^2 + \cgp^2 \sigma_\eta^2 +2\cgp {\rm cov} (\delta_L,\eta_\parallel)$. 
Since we already fixed the value of $\bgp\sigma_g$, this provides us with the value of $\bgp$.
We can then generate spectra using this $P_s(k)$ and FGPA (Eq.~\ref{eq:FGPA}) and compute the resulting $\Ponedf$. In each bin in $k$, $P_s$ at iteration $n+1$ is defined from iteration $n$ as $P_s^{n+1}(k)=P_s^n(k) \times (P_{target}(k)/\Ponedf(k))$.
We stop after ten iterations, which provides a good agreement for $\Ponedf$.

However, the correlation function prediction (section \ref{sec:prediction}) for these selected values of the mock parameters implies a bias that is a bit larger than our target. 
This might be do to the fact that
the measured $\Ponedf$ includes the HCD contribution,
which increases the amplitude of the correlation function. 
We therefore make small adjustments of $\agp$ and $\bgp$ to get a better value of the bias and keep a good value of 
$\Fbar$, and perform again the previous procedure for $P_s(k)$. 

\begin{figure}[H] 
\centering
\includegraphics[scale=0.4]{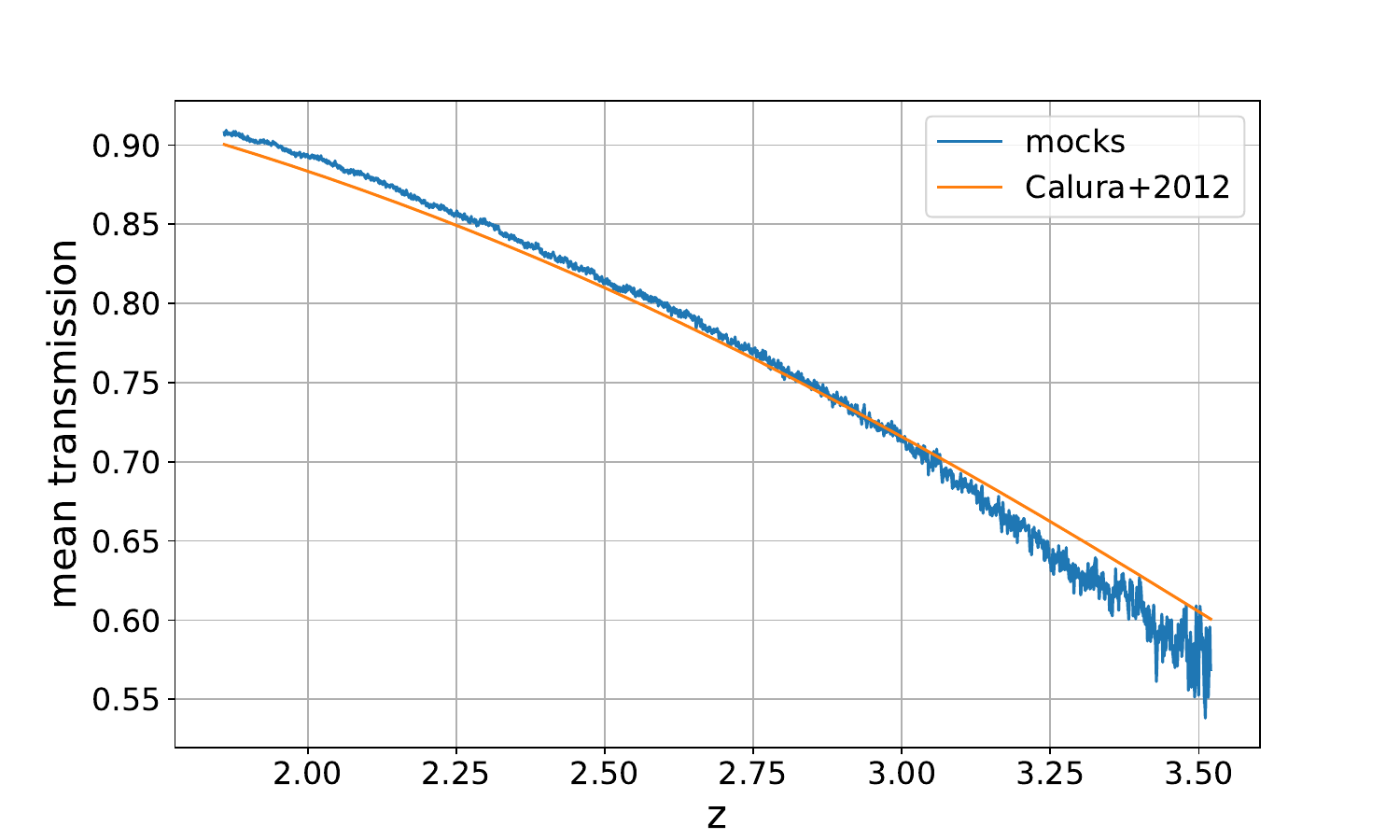} 
\caption{Mean transmitted flux fraction versus redshift, 
$\Fbar(z)$. 
The blue curve is $\Fbar(z)$ as measured on the mocks in the region $91.2<\lrf<120$ nm,
while the orange line is the parameterization of published data by Ref.~\cite{Calura+12}.
}
\label{fig:Fmean}
\end{figure}

\begin{figure}[H] 
\centering
\includegraphics[scale=0.6]{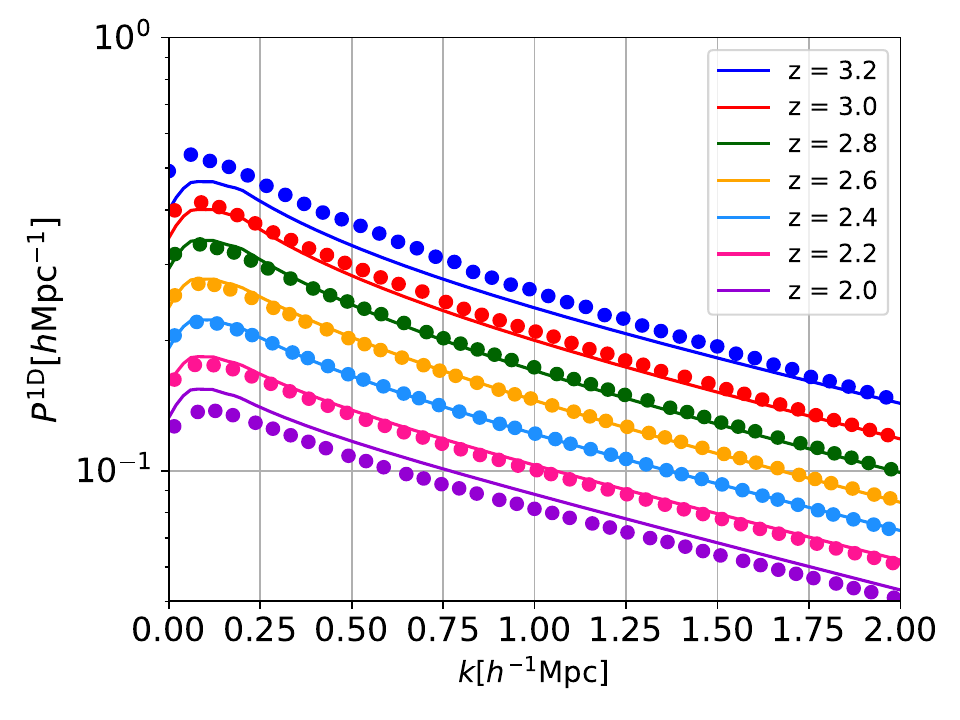} 
\caption{The measured 1d power spectrum, $\Ponedf$,  of the mock data in bins in redshift compared to the target 1d power spectrum.
}
\label{fig:P1Dmocks}
\end{figure}

After five iterations 
there is fair agreement for all parameters, as illustrated in Fig.~\ref{fig:Fmean} for $\overline{F}$,
in Fig.~\ref{fig:P1Dmocks} for $\Poned(k_\parallel)$,
and in  Figs.~\ref{fig:xi_lya} 
and \ref{fig:b_lya}
of Sect.~\ref{sec::analraw} for the bias parameters and overall correlation function.
For $\overline{F}$, the mocks have $\approx1\%$ excess transmission
at low redshift and a small excess at high redshift.
For $\Poned$, Fig. \ref{fig:P1Dmocks}  shows that the mocks have $\approx10\%$ too much power for $z>3.1$ and $\approx10\%$ too little for $z<2.1$.
This has little impact on the correlation functions because there are few forest pairs with $z>3.1$ and the flux fluctuations for $z<2.1$ are dominated by instrumental noise.
Figure~\ref{fig:b_lya} shows that the raw mocks reproduce the targeted bias parameters at the percent level.
However, Fig.~\ref{fig:xi_lya} shows noticeable differences in target and returned correlations for separations near the line of sight  ($\mu>0.95$). 
Finally, we note that the bias parameters found for the quasar-forest cross-correlations, as shown in Table \ref{table::best_fit_parameters_lyaonly} differ by of order 10\% from those found for the auto-correlations.

\subsection{High column density systems (HCD)}

The \lya\ absorption in the IGM  is responsible for most
forest absorption but the production of realistic mocks
requires the addition of other types of absorption 
in the post-production stages described in \cite{Herrera_2024}.
One type is absorption due to high column density systems (HCDs),
i.e. condensed systems with neutral-hydrogen column-density $N(\Hone) > 10^{17.2}\mathrm{cm}^{-2}$. HCDs with column density $>10^{20.3}\mathrm{cm}^{-2}$
are called damped \lya\ systems (DLAs) and can be identified in 
eBOSS or DESI spectra of moderate signal to noise.
About 6\% of such spectra contain a DLA\cite{DLAcatalogdr16}.

Like IGM absorption, absorption by HCDs is dominated by \lya\ absorption but
is spread out along the line of sight because of the system's
absorption profile.
This modifies the flux correlation function by suppressing
high-$k$ modes along the line of sight \cite{Rogers_2018}.
For instance, a DLA of column density of $10^{20.3}\mathrm{cm}^{-2}$ will 
entirely absorb the flux over a wavelength range corresponding
to a comoving distance of $\approx5  h^{-1}$Mpc.
The modeling of this effect in eBOSS and DESI spectra will
be studied in a forthcoming publication [Tan et al, in preparation].
The eBOSS and DESI analysis pipelines mask the highly absorbed
regions around detected DLAs so one needs only model
the effect of HCDs that are undetected, either because their column
density is too small ($<10^{20.3}\mathrm{cm}^{-2}$) or because they
are in low signal-to-noise forests.
The current modeling of HCDs in DESI involves two parameters, one giving the strength of HCD absorption and a second giving the 
length scale of the most important absorbers.

In order to include HCDs in the second stage of mock production,
the first stage produces
a catalog of possible HCD positions is selected from
peaks of the large-scale density field, $\delta_L$, with  the threshold set so as to produce a redshift-independent HCD bias $b_{\rm HCD}(z)=2$ \citep{DLAbias+18}.
The threshold is set
using an analytic evaluation  of eqn. A2 
in appendix A of Ref.~\cite{FontMiralda12}. The selected peaks are then  Poisson sampled to follow the HCD redshift distribution of the default model from the IGM physics package pyigm\footnote{https://github.com/pyigm/pyigm}~\cite{Prochaska+17}, which is fitted to a selection of literature results, summarised in table 1 of Ref.~\cite{Prochaska+14}.
Then a column density is randomly ascribed to each HCD to follow the column density distribution of the same model. 
The radial velocity of the HCD is obtained from the three velocity-component boxes, as for the quasars.

Our implementation of HCDs is clearly approximate because
there are no measurements of the number or bias of HCDs in the
range $10^{18}<N(\Hone)<10^{20}{\mathrm cm^{-2}}$.
We can expect, however, that the most important HCD effects
on the correlation function come from the unmasked HCDs with
the highest column density, near $N(\Hone)\approx10^{20}$
where the extrapolation of the measurements should be reliable.

\subsection{Metals}
The other major secondary source absorption is that due to non-\Hone\ species, i.e. metals, either in the IGM or in
condensed systems.  The presence of such absorption complicates the modeling of the \lya\ forest correlations because the transformation of wavelengths to comoving radial distances necessarily assumes that features are due to \lya\ absorption.
The presence of metallic absorption and its correlation with \lya\ absorption results in "phantom" correlations superimposed on the physical correlation function but shifted in the radial direction.
This effect is simple to model as described in Sect. 4 of \cite{dMdB+20}.
Because of their proximity to the \lya\ wavelength, the most important metallic species are four silicon transitions:
SiII(1190), SiII(1193), SiII(1260), and SiIII(1207).

The procedure for adding metallic absorption in the post-production stage is described in \cite{Herrera_2024}.
The currently used process assumes that absorption by metals are proportional
to that by HI and follows the same procedure as
the eBOSS mocks described in \citet{Farr+19}.
An alternative procedure, proposed by \cite{Farr+19},
would be to place the metals in
density peaks as is done for quasars and HCDs.

\section{Implementation and production}
\label{sec:prod}

The code was run at  National Energy Research Scientific
Computing Center (NERSC)  
on Cori machines without requiring the use of different nodes in parallel. The nodes of this machine have 64 threads and 128 Gb memory, which limits the number of voxels of the boxes. 
We have used boxes with $2560\times2560\times1536$ cubic voxels of $2.19 h^{-1}$Mpc side. 
 One such box is not enough to cover neither the north galactic cap (NGC) nor the south one (SGC) of the eBOSS or DESI footprints and we use five boxes for the NGC and two boxes for the SGC as illustrated by Fig.~\ref{fig:chunks}. This means that there is no correlation between spectra produced in different chunks. However if we consider separation below 200 $h^{-1}$ Mpc, the proportion of pixel pairs with pixels that belong to different chunks is negligible.

The code is written in python and is available on github\footnote{\verb+https://github.com/igmhub/SaclayMocks+}. 
For one realization of the mocks we must produce 7 independent chunks, which we finally merge. 
The procedure consists of four  steps summarized in Table \ref{table:CPU}.
First, the module \verb/make_box.py/ produces the 13 boxes for the considered chunk (1 density box, 3 lognormal density boxes, 3 velocity boxes and 6 velocity gradient boxes).  This module was run using one Cori node, taking advantage of the pyFFTW\footnote{\verb+https://github.com/pyFFTW/pyFFTW+ based on FFTW library, \verb+http ://www.fftw.org+} package to run the FFT in parallel on the 64 threads of the node. The module \verb+draw_qso.py+ then samples the box to draw quasars, as described in section~\ref{sec:qso} and the module \verb+make_spectra.py+ computes the density along each line-of-sight. These two modules run on 512 threads over 16 nodes. 
The box is split into 512 slices along a direction perpendicular to the line-of-sight and each thread treats one slice. 
This constitutes the ``pre-production'' of the realization. This is where we have strong constraints in terms of CPU consumption and, in the case of \verb/make_box.py/,  in terms of memory.

The second step is the ``post-production'', which involves combining the pieces of spectra from the different 
slices 
of the box, adding small-scale field, $\delta_S$, and applying the FGPA  (\verb+merge_spectra.py+).
This produces a set of "transmission files" containing
for each line-of-sight, $q$, the transmitted-flux fraction
$F_q(\lambda)$.
A catalog of HCD is also produced which, if desired, can
be added to the $F_q(\lambda)$ at a later stage.

After the production of the transmission files following
the steps of Table~\ref{table:CPU},
the final step of the mock production 
samples the set of transmission files to provide
the desired density of quasars ($\approx20{\rm deg^{-2}}$
for eBOSS up to $\approx100{\rm deg^{-2}}$ for DESI)
and then 
transforms the transmission-fractions $F_q(\lambda)$ 
into fluxes, $f_q(\lambda)$.
This is done 
with a script named \texttt{quickquasars}
\footnote{\verb+https://github.com/desihub/desisim/blob/main/py/desisim/scripts/quickquasars.py+}, 
which is a compilation of DESI code within the
\verb+desisim+ package\footnote{\verb+https://github.com/desihub/desisim+}, 
as described in Section 2.2 of \cite{Herrera_2024}.
It multiplies the $F_q(\lambda)$ by quasar continua 
chosen from a library of realistic quasar spectra \cite{simqso:2021}.
Instrumental noise is also added at this stage.
Options  for \verb+desisim+ include multiplication
by Voigt profiles due to HCDs and the addition of absorption by metals. 
The produced spectra files in the same format as real data.

In what follows, the set of $F_q(\lambda)$
obtained before applying \verb+desisim+ 
are referred to as "raw" mocks.
The set of fluxes $f_q(\lambda)$ obtained by \verb+desisim+
are referred to as  "cooked" mocks. 

Table \ref{table:CPU} presents the CPU and memory requirements for the different modules of the production
of the transmission files. The first three lines are the main parts of the pre-production, which requires a total time of about 2500~h $\times$ threads of CPU.  
The last line is \verb+merge_spectra.py+, which dominates the post-production and requires about 350  h $\times$ threads, and does not have significant memory requirements. So the post-production can be run easily.

\begin{table*}
    \caption{ CPU and memory requirements for mock production. 
    Columns 2, 3 and 4 give requirements
    for one chunk and column 5 gives the requirement
    for a complete realization of 7 chunks on Cori.
    }
    \label{table:CPU}
    \begin{tabular}{l | r | r | r | r }
    & \multicolumn{3}{c|}{1 chunk} &  1 realization 
    \\
 & time (min) & node (64 threads) & memory (Gb) & time $\times$ thread (h)\\
    \hline
\verb/make_box.py/ & 210 & 1 & 120 & 1570  \\
\verb/draw_qso.py/ & 2 & 16 & not critical & 240  \\
\verb/make_spectra.py/ & 6 & 16 & not critical & 720  \\ \hline
\verb/merge_spectra.py/ & 3 & 16 & not critical & 360  \\
         \end{tabular}
\end{table*}

A hundred sets of transmission files have been produced. The output of the pre-production are saved 
so that if we decide to change the parameters of the mocks that enter eq.~\ref{eq:FGPA}, we only have to re-run the post-production, which does not require important memory or CPU resources.

\section{Analysis of the mock data sets}

The mock data sets consist of a set of \lya~forests and sets of two types of discrete objects,
quasars and HCDs.  
To verify that the sets have the expected characteristics,
we studied three types of correlations:
forest-forest auto-correlations;
forest-quasar  cross correlations; 
and quasar-quasar auto-correlations.
(Correlations with HCDs will be studied in [Tan et al, in preparation])
In this section, we first describe the analysis procedure and then
present the correlation functions measured 
for raw and cooked mocks with 
eBOSS quasar densities ($\approx20{\rm \,deg^{-2}}$) and noise.
We then give the results of fits of the correlations 
to a \lcdm-based model, as summarized in 
Table \ref{table::best_fit_parameters_lyaonly}.

\subsection{Analysis procedure of forest correlations}
\label{sec::analprocedure}

Correlations involving the forests were analyzed using the standard eBOSS and DESI pipeline, picca\footnote{\verb+https://github.com/igmhub/picca+}, which is described for instance in Ref.~\cite{dMdB+20},. 
The main steps are as follows:
 

\begin{enumerate}
\item 
Calculation of  
the flux-transmission field, $\delta_q(\lambda)$, 
for each quasar, $q$.
The calculations uses
the flux $f_q(\lambda)$ (for the cooked mocks), 
or the transmission, $F_q(\lambda)$ (for the raw mocks):
\begin{equation}
    \delta_{q}(\lambda) =
    \frac{
    f_{q}(\lambda)
    }{
    \overline{F}(\lambda)C_{q}(\lambda)
    } - 1. \hspace*{10mm}
    {\rm or} \hspace*{5mm}
    \delta_{q}(\lambda) =
    \frac{
    F_{q}(\lambda)
    }{
    \overline{F}(\lambda)    }
    -1 \; .
    \label{equation::definition_delta}
\end{equation}
Here $\overline{F}(\lambda)$ is the mean transmitted flux fraction 
and $C_q(\lambda)$ is the estimated
unabsorbed quasar continuum.  
We calculate the $\delta_q(\lambda)$ 
in a wavelength range between the \lya~and Ly$\beta$ quasar lines:
$1040<\lrf<1200$\AA.

In the standard analysis 
of the cooked mocks or of data
(Sect. 2.4 of \cite{dMdB+20}) 
the product 
$\overline{F}(\lambda)C_{q}(\lambda)$ 
is determined by fitting each forest
to the mean
forest spectrum modified by two free parameters per quasar reflecting the
quasar brightness and spectral index.
Such a simple procedure does not fully take into account the peculiar features of individual quasar spectra.
Fortunately, the sources of quasar spectral diversity are believed to be due to  astrophysical conditions near the quasar and are therefore unlikely to lead to correlated absorption in forests separated by cosmological distances.

For the raw analysis, the mean transmission
$\overline{F}(\lambda)$ (Fig.~\ref{fig:Fmean})
can be estimated by averaging
the transmission $F_q(\lambda)$ over all forests, $q$,
in small wavelength ranges.
However, because these $F_q$ are all near a host quasar,
they are not a representative sample of the whole IGM,
and the correlation with the host results in a slight 
underestimation of $\overline{F}$ of $\approx10^{-4}$.
(see Sect. \ref{sec::analraw}).  
An unbiased value 
$\overline{F}(\lambda)$ 
could be  measured on lines-of-sight that were
randomly selected rather than being required to be
in front of quasars.  
Lacking such data, we added a constant to the $\overline{F}$
so that the mean $\delta_q(\lambda)$ far in front of
the quasars had the expected value of the cross-correlation
at $r\approx200\hmpc$.
(See Fig. \ref{fig:hostxcf})
This small shift has negligible effect on the best-fit model
parameters.

\item Use of the  fiducial cosmological model 
used to construct the mocks to transform forest-pixel pair
separations or quasar-pixel separations 
$(\Delta z,\Delta\theta)$ into co-moving distances $(\rperp,\rpar)$
(Sect. 3.1 of \cite{dMdB+20})

\item Calculation of the forest auto correlation function
and quasar-forest cross correlation 
in bins
$\Delta\rperp=\Delta\rpar=4\hmpc$
(Sections 3.2 and 3.3 of \cite{dMdB+20}):
\begin{equation}
	\xi_{A}^{auto} = \frac{
    \sum\limits_{(i,j) \in A} w_{i}w_{j} \, \delta_{i}\delta_{j}
    }{
	\sum\limits_{(i,j) \in A} w_{i}w_{j}
    }
    \hspace*{10mm}
    \xi_{A}^{cross} = \frac{
    \sum\limits_{(i,q) \in A} w_q w_{i} \, \delta_{i}
    }{
	\sum\limits_{(i,q) \in A} w_q w_{i}
    }.
    \label{equation::xi_estimators]}
\end{equation}
For the auto-correlation,
the sum is over pixel pairs in the $(\rperp,\rpar)$ bin $A$ and
the $w_i$ are weights chosen to optimize the measurement.
For the cross correlation, the sum is over quasar
pixel pairs.

\item  Calculation of the covariance matrix of the 
auto- and cross-correlations by subsampling
(Sections 3.2 and 3.3 of \cite{dMdB+20}).
The method consists of measuring the correlation
functions in different regions of the sky
and deducing the covariances from the variations
between different regions.

\item Fit of the measured $\xi$ with a 
\lcdm-based model
(Sect. 4 of \cite{dMdB+20}).
\end{enumerate}

The model used to fit the correlations
is based on  the \lcdm~power spectrum modified
to allow for a change in the position of the BAO peak with respect to the fiducial model
and by Kaiser factors to describe biasing and RSD.   

The parameters of the fits are listed in Table~\ref{table::best_fit_parameters_lyaonly}..
The two parameters $(\aperp,\apar)$ describe the position of the BAO peak.
The density and velocity biases, $(\blya,\biasetalya)$ are related by the
fit parameter $\betalya=f\biasetalya/\blya$,
 where $f\approx1$ is the growth rate at $z\approx2.3$.

Fits for the quasar-forest cross-correlation include 
a parameter, $\Delta r_\parallel$, that, for real
data, compensates for
systematic errors in quasar redshift estimation.
We include it in the fit of the mock data, yielding,
as expected, a value consistent with zero.
The values of $b_{QSO}$ and $\beta_{QSO}$ for these
fits are fixed to the values determined for the quasar
auto-correlation (Sect.~\ref{sec::analqso}).

For cooked mocks and for data,
the model correlation and cross-correlations functions
must be multiplied by 
a ``distortion matrix"  which corrects
for the effect of the continuum fitting used in 
step 1 for the cooked mocks
(Sect. 3.5 of \cite{dMdB+20}).

\subsection{Forest correlations in the raw mocks}
\label{sec::analraw}

The raw mocks were analyzed using the second form of eqn.~\ref{equation::definition_delta}
to define the flux-transmission field.
The aim was first to verify that the produced forest-forest auto-correlations corresponded well to
the expectations defined in Sec. \ref{sec:prediction}.
Second, we investigated the extent to which the forest-quasar cross-correlation was correctly
predicted by the same model as that for the auto-correlation.

The \lya\ auto-correlation was computed for 30 raw mock realizations
of the eBOSS footprint and quasar density.
The correlations were calculated for the entirety
of the data, corresponding to a mean pixel-pair redshift
$z_{eff}=2.261$.
The correlations were also calculated in 
four redshift bins of similar statistics.
Following \cite{2019deSainteAgatheDR14} redshift bins were
defined by the mean quasar redshift of all pairs of
forests.
Ranges for the mean quasar redshift were defined: 
$2.0<z<2.35$; $2.35<z<2.65$; $2.65<z<3.05$ and $z>3.05$. 
All forest-pixel pairs of the forest pairs were used to calculate the
auto-correlation for that range.
This led to measurements for the  $z_{eff}$ values
2.101, 2.237, 2.542, and 2.866.

\begin{figure}[H] 
\centering
\includegraphics[width=0.95\textwidth]{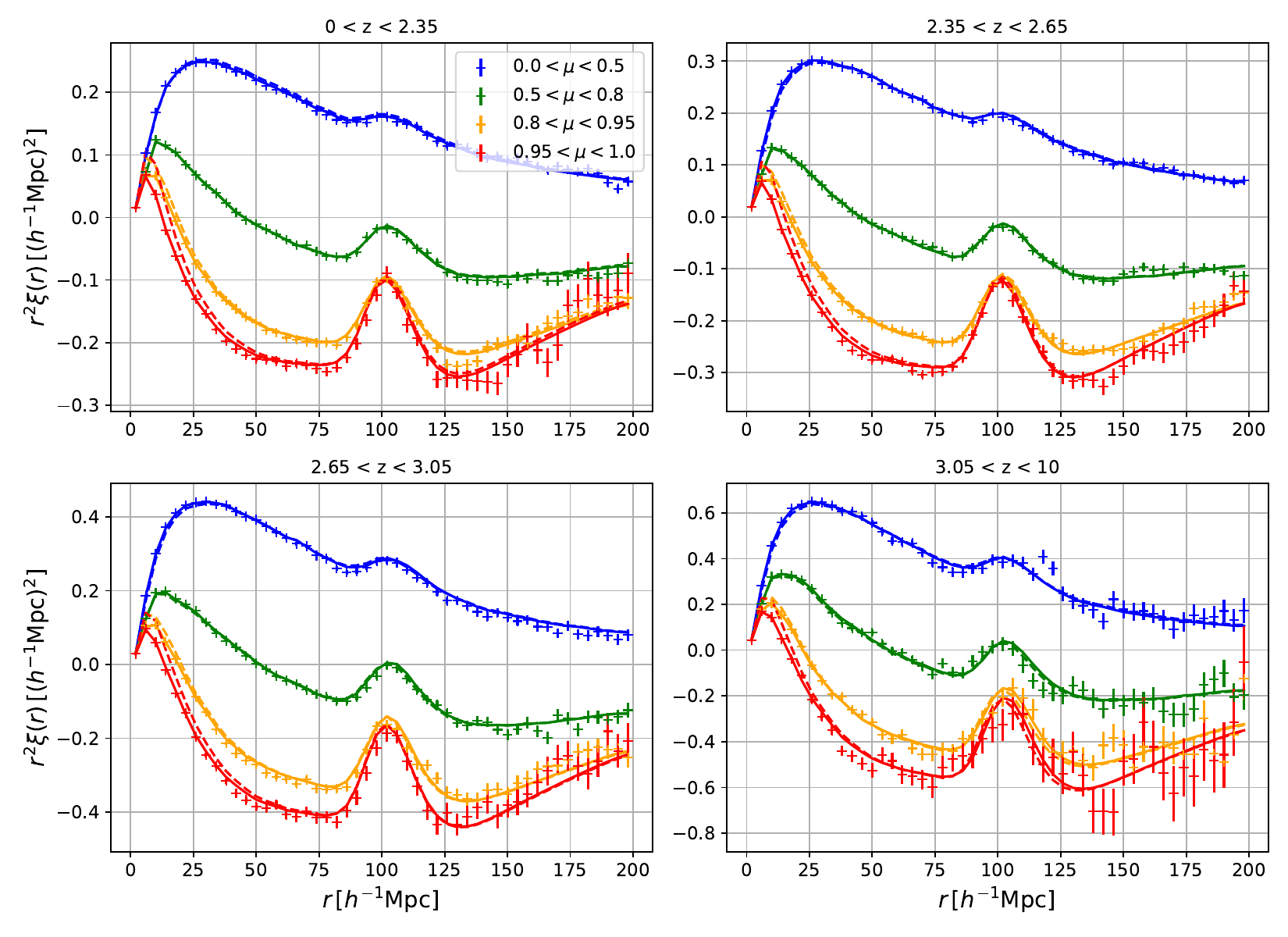} 
\caption{\lya\ auto-correlation of  30 raw mock realizations. The four panels correspond to different bins in redshift. Each one gives the average of the correlation functions calculated for the 30 realizations. The correlation functions are presented in four bins in $\mu$. The continuous lines give the fit by picca model and the dashed lines the prediction (section \ref{sec:prediction}).
}  
\label{fig:xi_lya}
\end{figure}

The results of fits to the correlations  
for the stack of 30 mocks
are presented in Fig.~\ref{fig:xi_lya}. 
In each redshift bin the correlation function is shown in 4 bins in $\mu=\rpar/r$. The dashed curve is the prediction (section \ref{sec:prediction}), which is fairly good, with only some disagreement at low $r$ in particular near the line-of-sight ($0.95<\mu<1$). 
The continuous line is the fit of the mock data with the picca model, it is in good agreement even at low $r$. 
The $\chi^2$ are 1557, 1608, 1628 and 1534 in the 4 bins in redshift for 1570 degrees of freedom.

\begin{figure}[H] 
\centering
\includegraphics[scale=0.48]{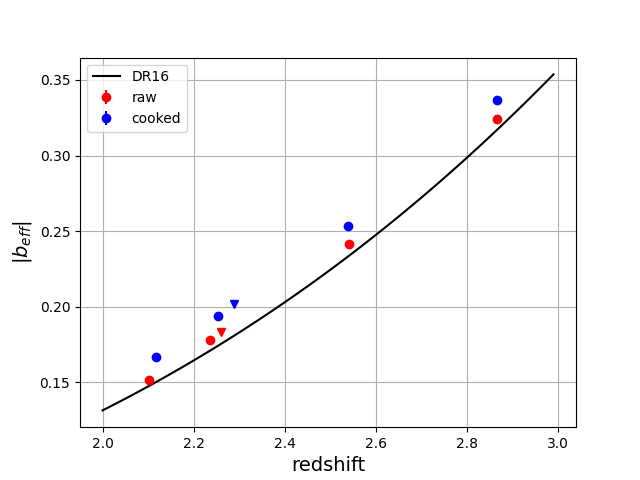}
\includegraphics[scale=0.48]{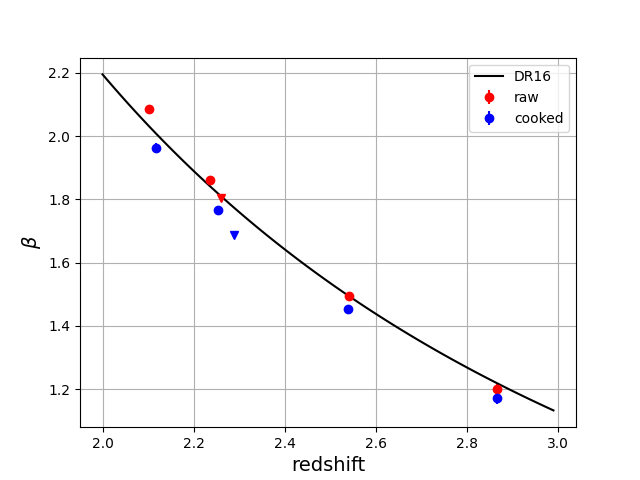}
\caption{
 \lya\ auto-correlation: the effective bias (left) and the RSD parameter (right) as a function of redshift. 
 The red points are for the raw mocks and the blue points for the cooked mocks.
 The triangles are for the fit over all redshifts (Table \ref{table::best_fit_parameters_lyaonly})
 The black lines are fits to the DR16 auto-correlation,
 that were used as the targets of the mocks
 }
\label{fig:b_lya}
\end{figure}

In Fig.~\ref{fig:b_lya} the red points show $b_{{\rm eff},\lyma}$ and $\beta_{\lyma}$ measured  in the raw mocks in four redshift bins 
for the average of 30 raw mock realizations. 
The figure also shows the same quantities 
for DR16 data together with power law fits to them, which were used as input parameterization for the mocks. 
We note that the values obtained for the raw mocks are consistent with those of the 
DR16 data. Due to the small error bars obtained on the mocks we can see a small disagreement with the fit in the case of $\beta_{\lyma}$ at low $z$. This level of disagreement is much lower than the error bar of the data.

We next studied
the  raw forest-quasar cross-correlations.
They were not used to tune the mocks and, as such, their measurement
in the mocks is a consistency check to determine to what extent the auto-correlation and cross-correlation can
be described by the same model.

We first studied the cross-correlations between a quasar
and its own forest, i.e., $\xi(\mu=1)$ given by 
the mean flux-transmission in front of a quasar.
This correlation cannot be measured in the 
standard analysis of the cooked mocks (nor in the data) because 
the correlation affects the forest flux in front
of the host quasar and is therefore  absorbed
into the continuum template. 
We can measure the correlation with the
raw mocks because the mean transmission,
$\overline{F}(z)$ can be determined,
as explained in Sect.~\ref{sec::analprocedure}, by
averaging the $F_q(\lambda)$ over all lines-of-sight
and then correcting for the slight underestimation
caused by the correlation with the host quasar.

\begin{figure}[H] 
\includegraphics[scale=0.48]{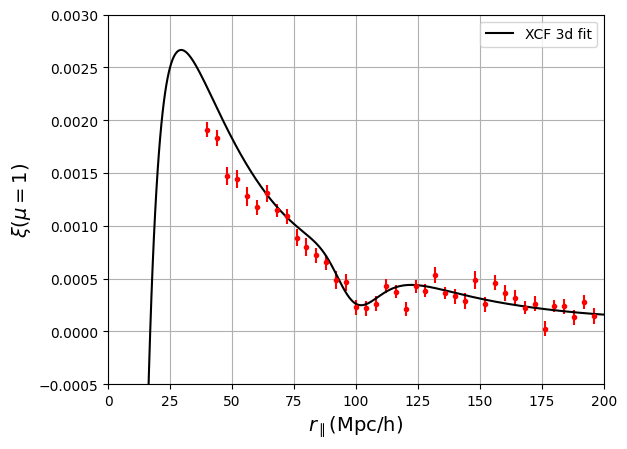}\includegraphics[scale=0.48]{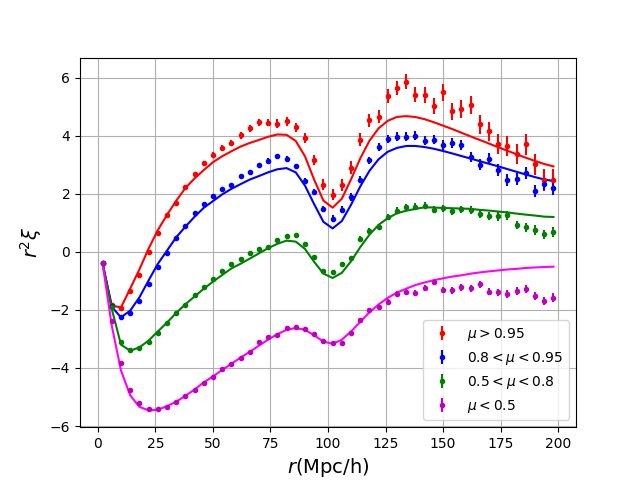}
\caption{Raw cross-correlation between host quasars and their forest (left) and the raw 3d cross-correlation in different $\mu$ ranges (right).
The solid lines show that best fits to
the raw 3d cross-correlation 
(Table~\ref{table::best_fit_parameters_lyaonly}) 
}
\label{fig:hostxcf}
\end{figure}

The $\mu=1$ cross correlation is shown in the left plot of 
Fig. \ref{fig:hostxcf}.
The correlation is $\approx2\times10^{-3}$ at a separation
of $40\hmpc$ corresponding to the point in the forest nearest
the host quasar.
The correlation then drops off to $\approx10^{-4}$
far in front of the quasar.
The mean of $\delta_q(\lambda)$ in the forest is $\approx 2\times10^{-4}$.

\begin{figure}[H] 
\centering
\includegraphics[scale=0.48]{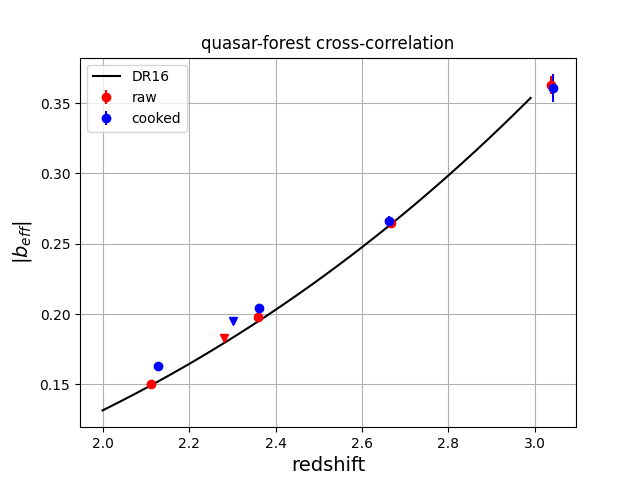}
\includegraphics[scale=0.48]{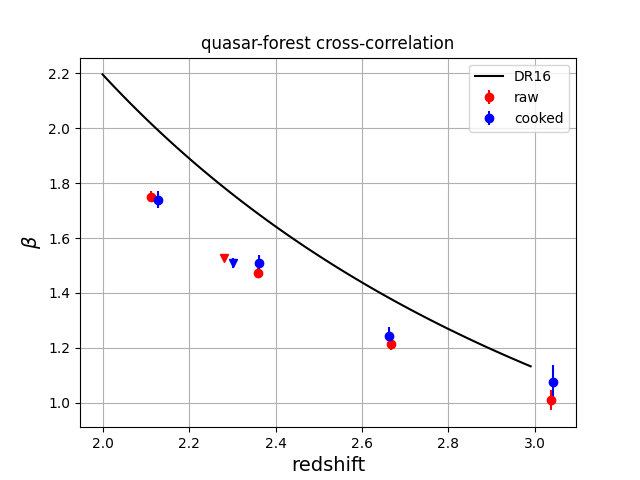}
\caption{
Same as Fig. \ref{fig:b_lya} except now for the
 quasar-\lya\ cross-correlation: the effective bias (left) and the RSD parameter (right) as a function of redshift,
 calculating assuming the quasar bias of \citep{Laurent+17}.
 The red points are for the raw mocks and the blue points for the cooked mocks.
 The triangles are for the fit over all redshifts (Table \ref{table::best_fit_parameters_lyaonly})
 The black line is a fit to the DR16 auto-correlation.
 }  
\label{fig:xb_lya}
\end{figure}

The $\mu\neq1$ cross-correlation 
in four ranges of $\mu$ is shown in the right plot of 
Fig. \ref{fig:hostxcf}.
The best-fit model shows non-negligible deviations from the measurements.
Figure \ref{fig:xb_lya} gives the best-fit bias parameters
as a function of redshift.
We see that the effective bias is in good agreement with
the expected value at all redshift while $\beta$ is about
10\% lower than expectations.

The $\mu=1$ cross-correlation shown in the left panel of 
Figure \ref{fig:hostxcf} indicates to what extent
the forests in front of a quasar do not represent
a random sample of the IGM.  
By definition, the mean $\delta$ in the IGM vanishes, whereas
the mean $\delta$ in the observed forest $\approx 2\times10^{-4}$.
This has little effect on the auto correlations (order $\delta^2$) but means that
extraction of the cross correlation in the raw mocks is significantly biased wherever the correlation
is less than $10^{-3}$. 
It is then not surprising that the fits of the raw mocks 
give cross-correlation values of  bias parameters
(Fig. \ref{fig:xb_lya})
that agree with expectations only at the 10\% level.
Fortunately, this issue has no observable effect on the
BAO parameters shown in Table \ref{table::best_fit_parameters_lyaonly}.

\begin{table*}
    \caption{    Best-fit parameters for stacks of 30 eBOSS mocks over the entire mock redshift range.
    Results are shown for the raw and cooked
    auto-correlation function (CF) and cross-correlation
    function (XCF).
    The first four rows show the best-fit $\chi^2$ for
    the fit with $N_{data}$ ($\rperp,\rpar$) bins with $N_{par}$ free parameters, leading to a $\chi^2$ probability $P$.
    The next rows give the best-fit parameters values for $(\apar,\aperp,b_\eta,\beta)$ and $\Delta z_{\parallel}$ for the XCF, and the derived parameters $b_{eff}$.
    The target values used to tune the mocks are shown
    for $\beta(z_{eff})$ and $b_{eff}(z_{eff})$ for the mean redshift of the measurement, $z_{eff}$
}
      \centering
    \scalebox{0.90}{
    \begin{tabular}{l | l | l | l | l}
& raw CF & cooked CF & raw XCF  & cooked XCF\\
    \hline
    $\chi^2$ &  1676.07  &  1562.54 &  3454.94  &  3502.09   \\
 $N_{data}$ &  1574   &  1574 &  3148  &  3148 \\
 $N_{par}$ &  4  &  4  &  5  &  5  \\
 $P$ &  0.031  &  0.548  &  0.0  &  0.0    \\
 $\alpha_\parallel$ &  1.001 $\pm$ 0.003  &  1.003 $\pm$ 0.005 &  1.002 $\pm$ 0.004  &  0.999 $\pm$ 0.005   \\
 $\alpha_\perp$ &  0.995 $\pm$ 0.004  &  0.995 $\pm$ 0.008 &  1.000 $\pm$ 0.004  &  1.0 $\pm$ 0.006  \\
 $-b_\eta$ &  0.196 $\pm$ 0.0003 &  0.2077 $\pm$ 0.0005 &  0.1754 $\pm$ 0.0006  &  0.1902 $\pm$ 0.001 \\
 $\beta$ &  1.804 $\pm$ 0.007   &  1.687 $\pm$ 0.008 &  1.583 $\pm$ 0.011  &  1.564 $\pm$ 0.015\\
 (target) & 1.809 & 1.774 & 1.782 & 1.758 \\
 $\Delta r_\parallel$ ($h^{-1}$Mpc)&  &  & $0.013\pm0.026$  & $-0.033\pm 0.035$ \\
 $b_{eff}$ & $0.1835\pm0.0002$ & $0.2021\pm0.0003$ & $0.1772\pm0.0004$ & $0.1935\pm0.0006$\\
 (target) & 0.1756 & 0.1807 & 0.1795 & 0.1831 \\
 $z_{eff}$ & 2.261 & 2.288 & 2.282 & 2.301 \\ 
 \hline
        \end{tabular}
    }
    \label{table::best_fit_parameters_lyaonly}
\end{table*}

\subsection{Forest correlations in the cooked mocks}
\label{sec::analcooked}

\begin{figure}[H] 
\centering
\includegraphics[width=0.95\textwidth]{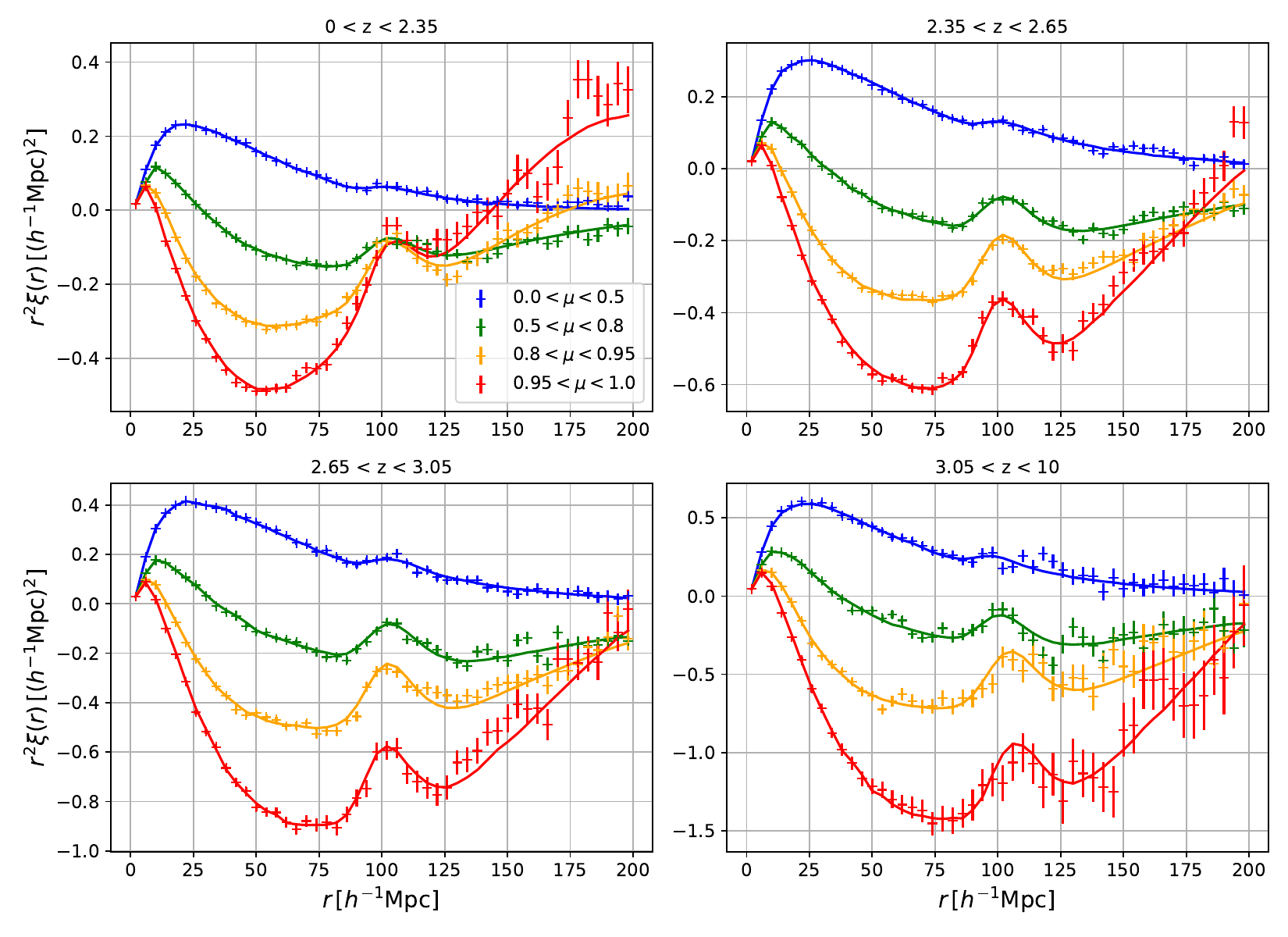} 
\caption{The same as Fig.~\ref{fig:xi_lya} except
now for the cooked mocks. Each graph gives the average of the correlation functions calculated in each redshift bins. The correlation functions are presented in four bins in $\mu$. The continuous lines are the fits with picca model.}  
\label{fig:cf}
\end{figure}

We now consider cooked mocks with continua and experimental noise. Fig.~\ref{fig:cf} presents the \lya\ x \lya\ correlation for the 30 realizations of cooked mocks. The analysis was performed in 4 bins in redshift and the results are presented in 4 bins in $\mu$. 
Comparing to Fig. \ref{fig:xi_lya} we note that the correlation function is considerably distorted by the fact that we have to fit the quasar continuum. This distortion is taken into account by the distortion matrix and the correlation function are rather well  fit over the range $20 < r < 180h^{-1}$ Mpc by picca model with $\chi^2$ of 1498, 1622, 1598, and 1629 in the four bins in redshift for 1570 degrees of freedom. 
The bias parameters, shown  as
the blue points in Fig. \ref{fig:b_lya} are in 
good agreements with expectations and with the
values derived for the raw mocks.

\begin{figure}[H] 
\centering
\includegraphics[width=0.95\textwidth]{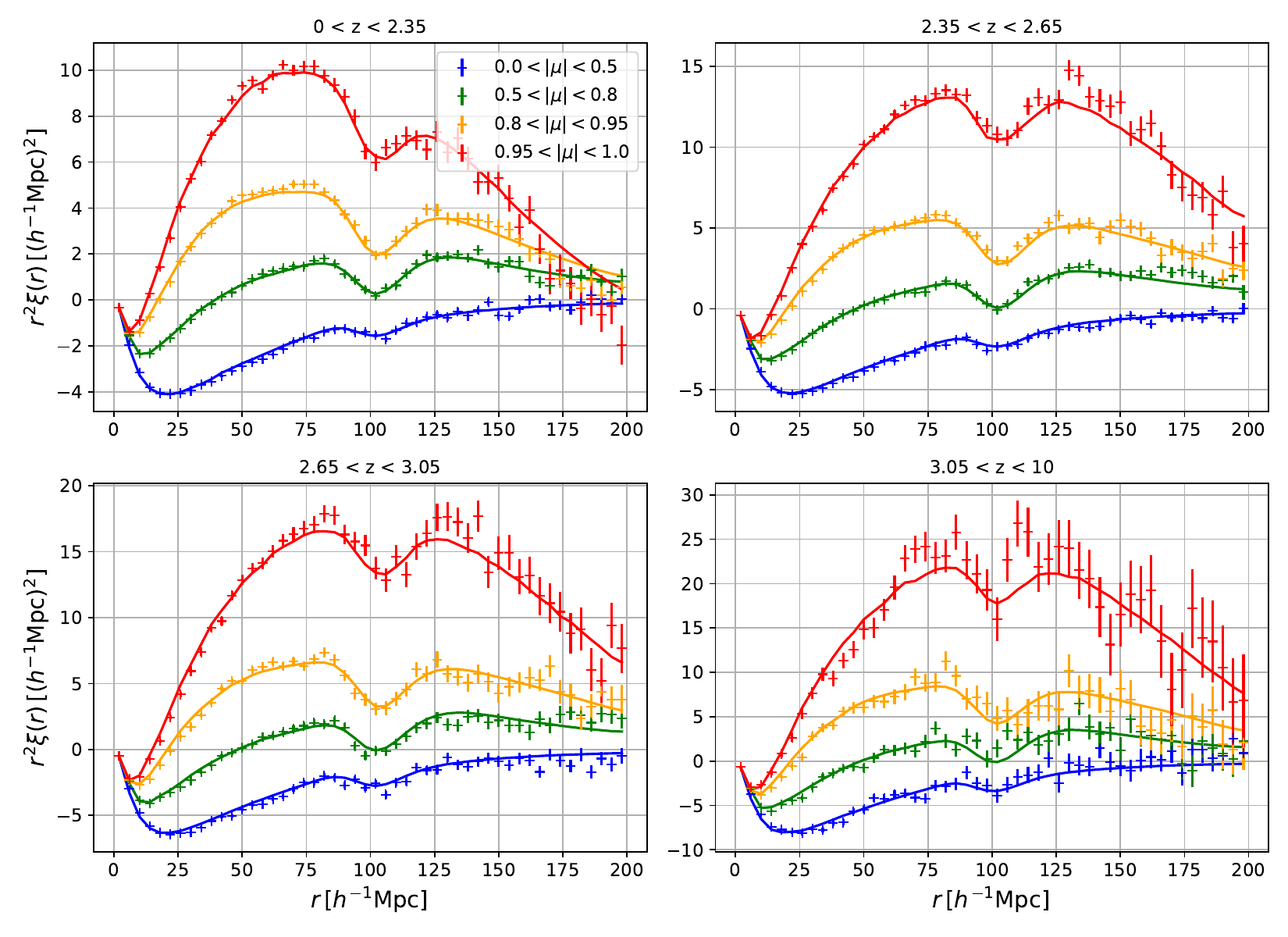} 
\caption{The \lya\ x QSO  cross-correlation for 30 realizations of 
cooked mocks. Each graph gives the average of the correlation functions calculated in each redshift bins. The correlation functions are presented in four bins in $\mu$. The continuous lines are the fits with picca model.}  
\label{fig:xcf}
\end{figure}

\begin{figure}[H] 
\centering
\includegraphics[scale=0.4]{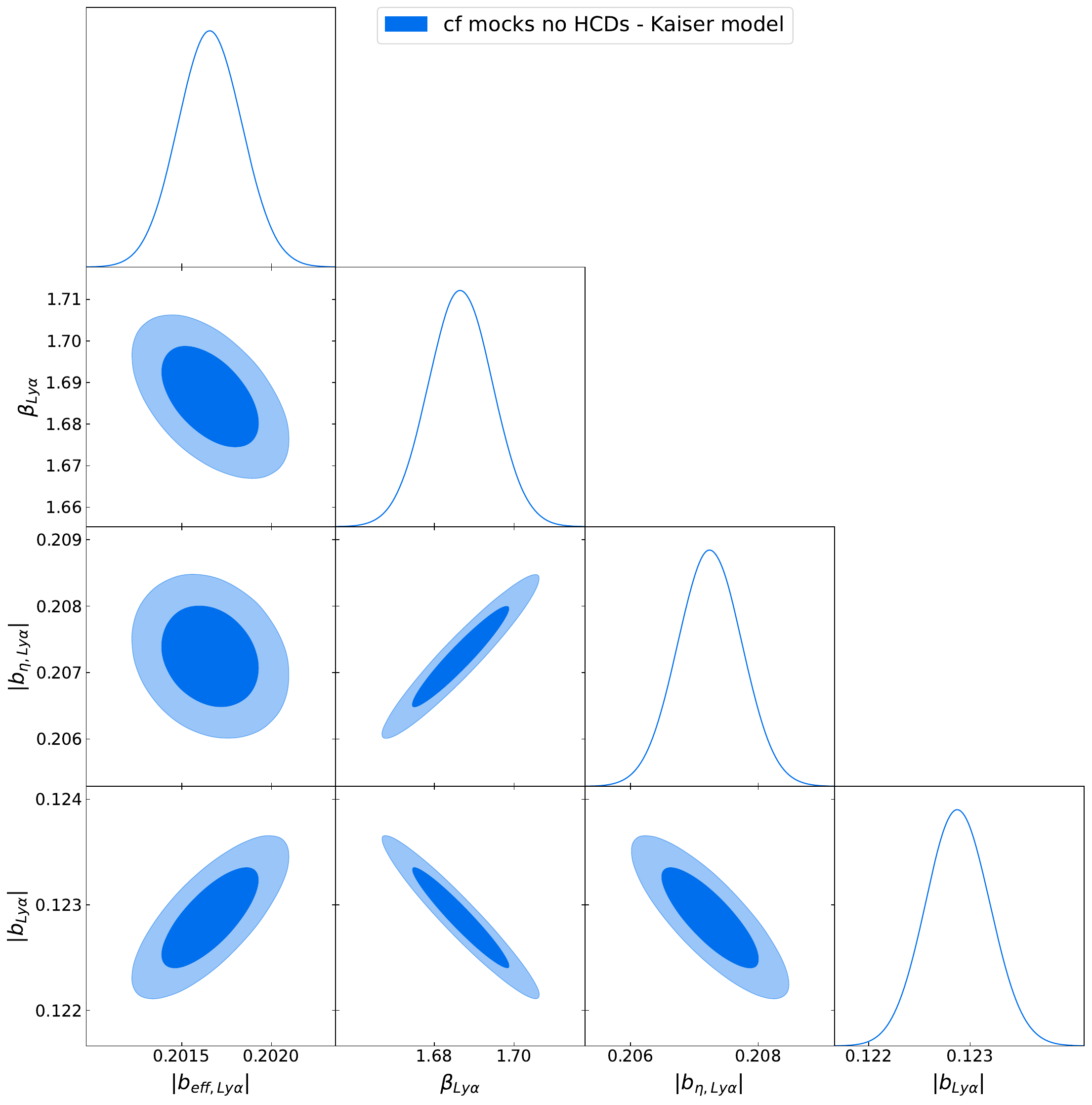}
\caption{
Correlations between the best-fit bias parameters of
the auto-correlations
 }  
\label{fig:biascorrelations}
\end{figure}

\begin{figure}[H] 
\centering
\includegraphics[scale=0.4]{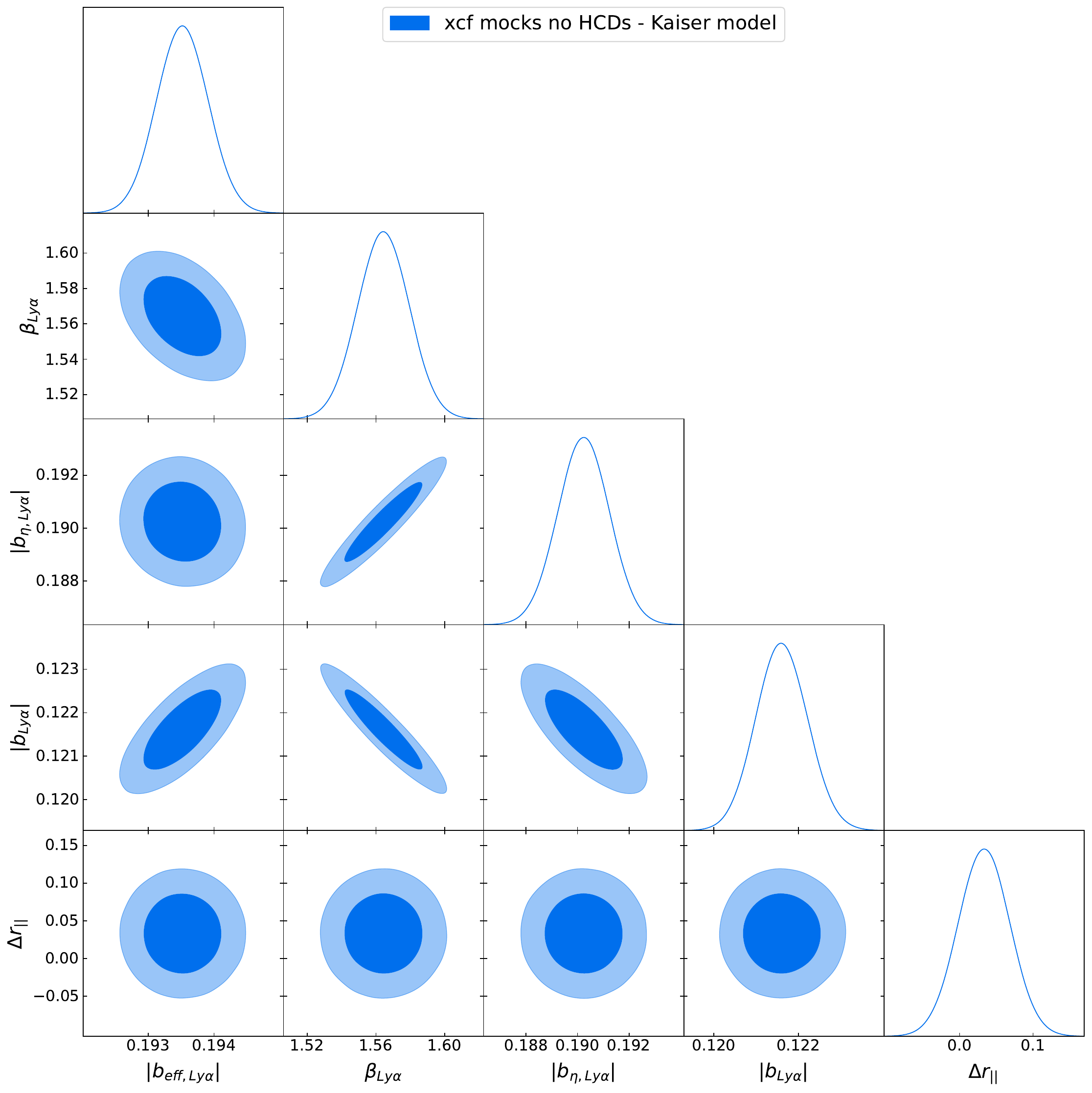}
\caption{
Same as Fig. \ref{fig:biascorrelations} but for
the cross-correlations
 }  
\label{fig:biasxcorrelations}
\end{figure}

\begin{figure}[H] 
\centering
\includegraphics[width=0.48\textwidth]{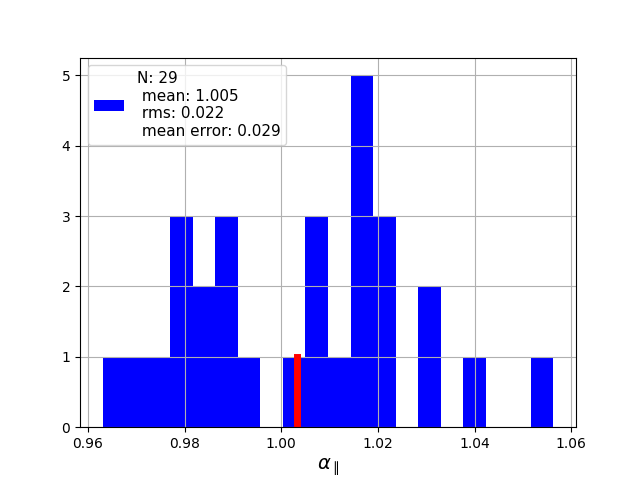} 
\includegraphics[width=0.48\textwidth]{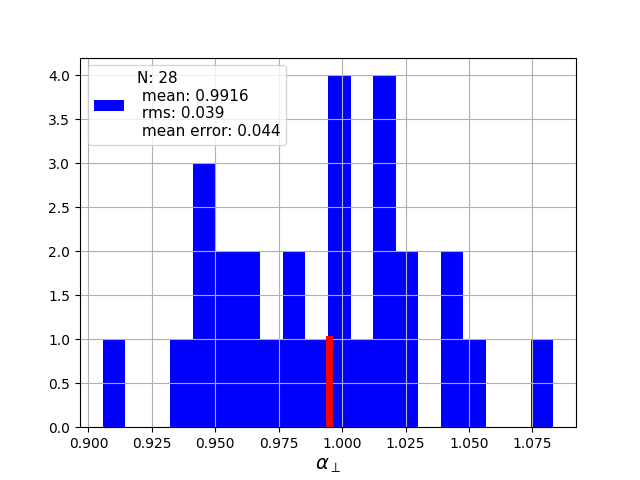} \\
\includegraphics[width=0.48\textwidth]{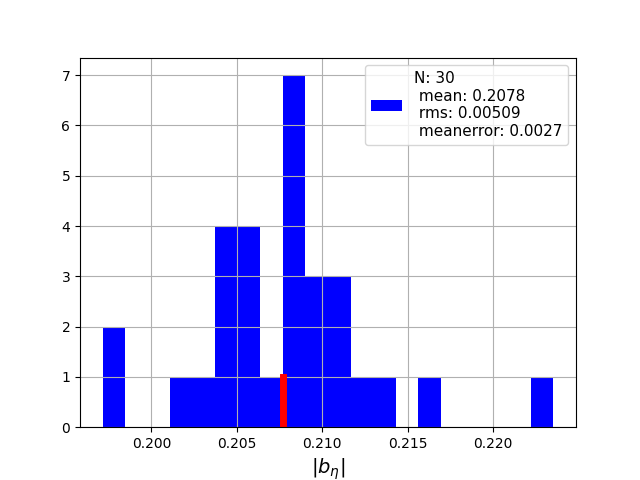} 
\includegraphics[width=0.48\textwidth]{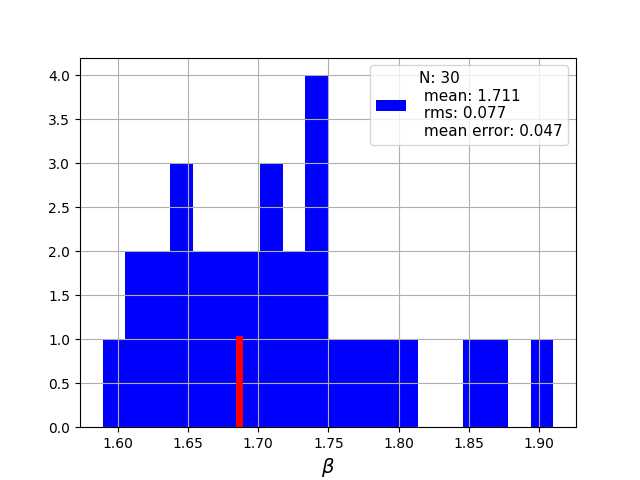} 
\caption{
The distributions of fit parameters for the auto-correlation for the 30 \lya-only mocks.
The red entry shows the fit value for the stack of 30 mocks.
The upper right corner lists the number of mocks included,
the mean and standard deviation of the distribution, and
the mean error reported by the fitter.
One  outlier is not included in the $\apar$ distribution
($\apar=1.50\pm0.12$)
and two outliers in the $\aperp$
distribution
($1.39\pm0.17$ and
$0.895\pm0.040$).
}  
\label{fig:pardists}
\end{figure}

\begin{figure}[H] 
\centering
\includegraphics[width=0.48\textwidth]{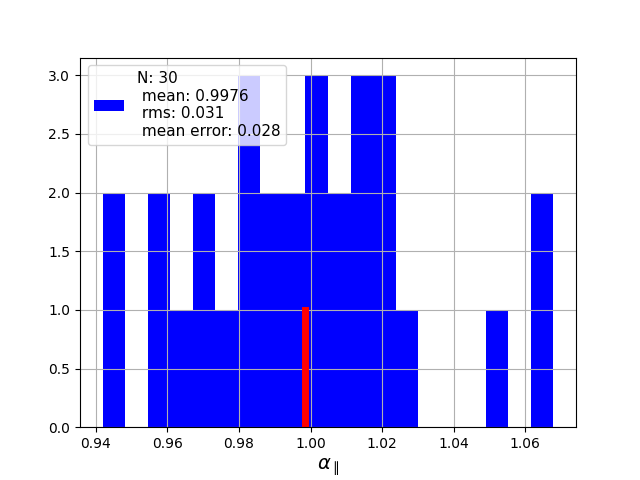} 
\includegraphics[width=0.48\textwidth]{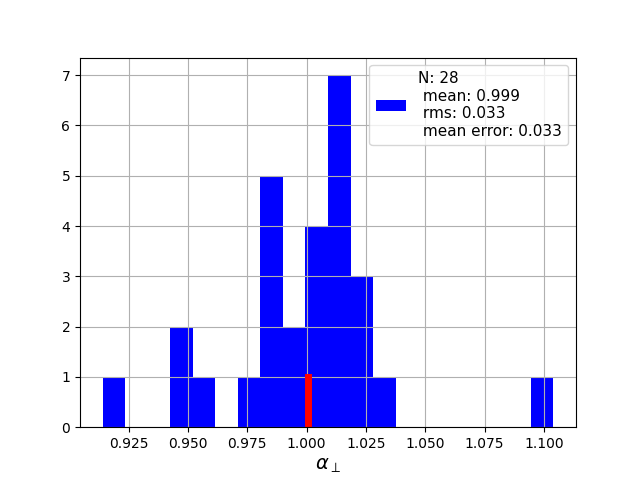} \\
\includegraphics[width=0.48\textwidth]{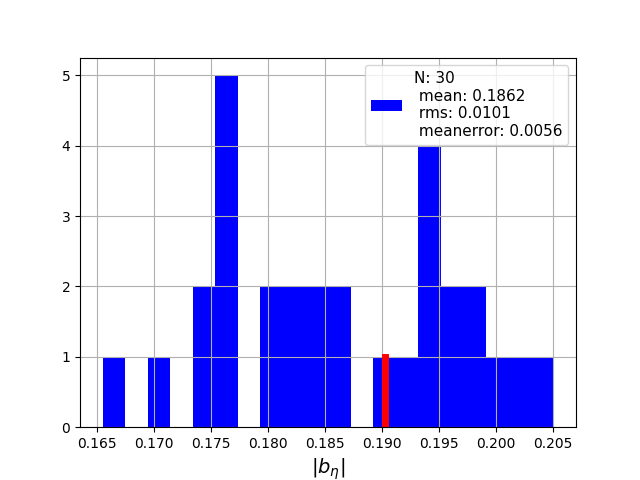} 
\includegraphics[width=0.48\textwidth]{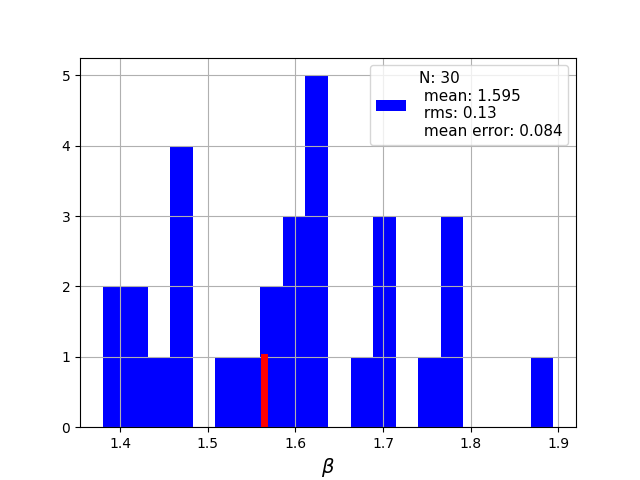} 
\caption{
Same as Fig. \ref{fig:pardists} except now for the cross-correlation.
Two outliers are not included
in the $\aperp$ distribution
($\aperp=1.31\pm0.08$ and
$\aperp=1.32\pm0.07$)
}
\label{fig:pardistsxcf}
\end{figure}

Fig.~\ref{fig:xcf} presents the quasar-\lya\ cross-correlation for a set of 30 realizations of cooked mocks with continuum and noise. 
As with the auto-correlation,
the analysis was performed in 4 bins in redshift and the results are presented in 4 bins in $\mu$. These cross correlations are rather well  fit over the range $20 < r < 180h^{-1}$ Mpc by the picca model with $\chi^2$ of 3233, 3410, 3228, 3410 in the four bins in redshift for 3143 degrees of freedom. 
The bias parameters shown in Fig. \ref{fig:biasxcorrelations}
are in agreement with those found for the raw mocks
but, again, the values of $\beta$ are about 10\%
lower than expected.

The correlations between the best-fit values of
the bias parameters are shown in Figs. \ref{fig:biascorrelations} and \ref{fig:biasxcorrelations}.
We see that for both the auto- and cross-correlations,
the most uncorrelated parameters are $b_{eff}$
and $b_\eta$.

In the standard BAO analysis~\citep{dMdB+20}, the primary purpose of the 
mocks is to verify that the analysis procedure returns unbiased
values of the model parameters and realistic uncertainties.
Figure \ref{fig:pardists} shows the distribution of
the best-fit parameters for the auto-correlation along with
the mean of the reported errors.  
We see that for the BAO
parameters, the mean returned values are both near
the expected values of unity, and the standard deviation of the distributions are consistent with the mean of the
reported uncertainties.
However, for $\aperp$, this is obtained only after eliminating
two outlier realizations with large errors
($1.39\pm0.17$ and
$0.895\pm0.040$).
For $\apar$, one realization was eliminated with $\apar=1.50\pm0.12$.
We conclude that for data samples of the eBOSS size the measurement of the BAO-peak position is sufficiently fragile to give significant deviations from expectations in of order 10\% of realizations. 
As first noted in \cite{dMdB+17}, this also results in slightly non-Gaussian errors for $\apar$ and $\aperp$.

For the two bias parameters, the reported uncertainties 
are nearly a factor two smaller
than the standard deviation of the distributions, implying
that the procedure significantly underestimates the uncertainties
in these parameters.
We note, however, that the errors on real data 
are likely to be dominated by
model uncertainties due
to effects not included in these mocks,
e.g. HCDs [Tan et al, in preparation].

Similar conclusions can be drawn for the cross-correlation
as illustrated in Fig. \ref{fig:pardistsxcf}.


\subsection{QSO auto-correlations}
\label{sec::analqso}

The quasar-quasar auto correlations
were measured using the standard techniques of comparing
the observed distribution of object pairs with random catalogs 
(Landy-Szalay statistic)
as described, for instance,
in \cite{Laurent+17}.
The auto-correlation was then fit with a \lcdm-based
model with bias parameters $b_{QSO}$ and 
$\beta_{QSO}=f b_{\eta QSO}/b_{QSO}$
where $f\approx1$ is the growth rate at $z\approx2.3$
and $b_{\eta QSO}\approx1$ is the velocity bias.
In the fits, the \lcdm~value of the growth rate was
not imposed so the fits yielded independent values
of $b_{QSO}$ and $\beta_{QSO}$.

\begin{figure}[H] 
\centering
\includegraphics[scale=0.35]{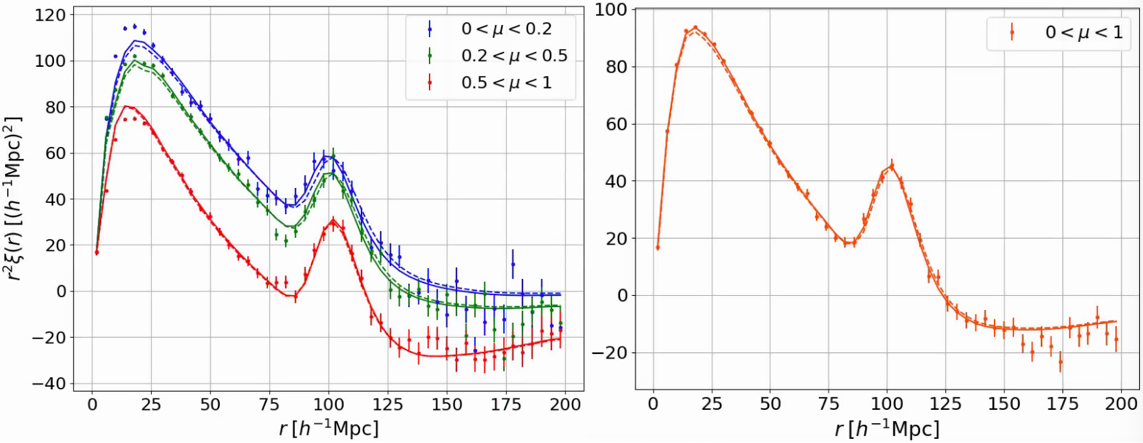} 
\caption{ The QSO auto-correlation of 10 mock realizations averaged over the four redshift bins. The left panel presents the correlation in three bins in $\mu$ and the right one the average over the whole range in $\mu$. The continuous lines give the fit by picca model and the dashed lines the prediction (Sect. \ref{sec:qso})}
\label{fig:xiQSO}
\end{figure}

Computing the auto-correlation of the quasars is rather CPU consuming and the mocks presented in this paper are not made to test the measurement of quasar auto-correlation. Therefore the auto-correlation was computed for only 10 realizations. The result is presented in Fig.~\ref{fig:xiQSO}. The right panel shows the correlation averaged over all bins in $\mu$, which is very close to the prediction (Sect. \ref{sec:qso}) , there is only a difference smaller than 2\% at $r<20 h^{-1}$ Mpc. The agreement is not as good in wedges in $\mu$, where the disagreement at small $r$ reaches up to 12\% (left panel). 

\begin{figure}[H] 
\centering
\includegraphics[scale=0.45]{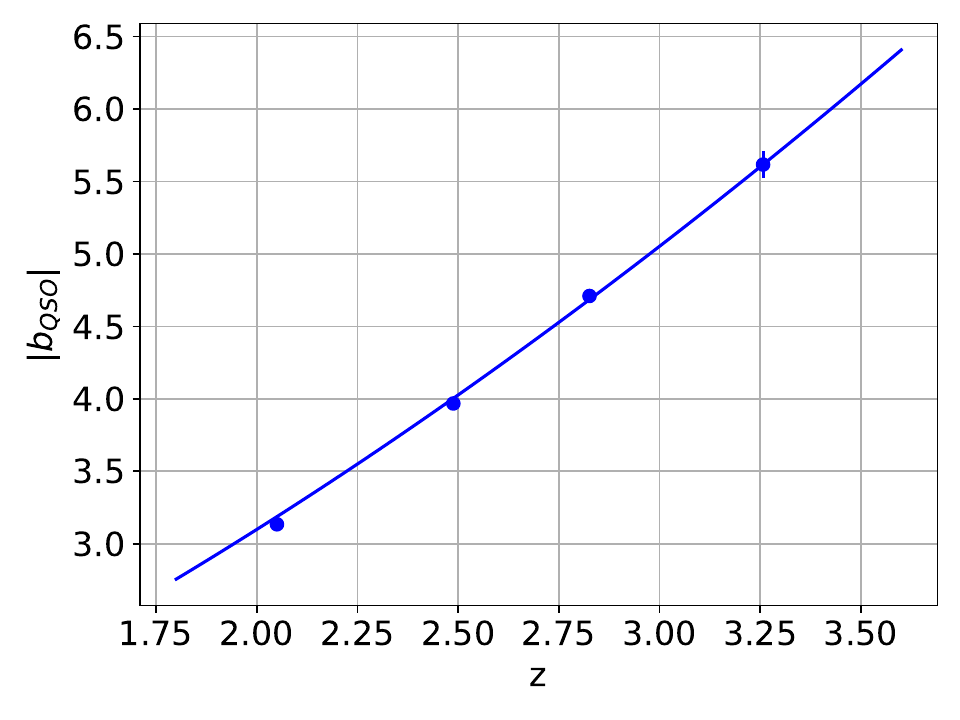} 
\includegraphics[scale=0.45]{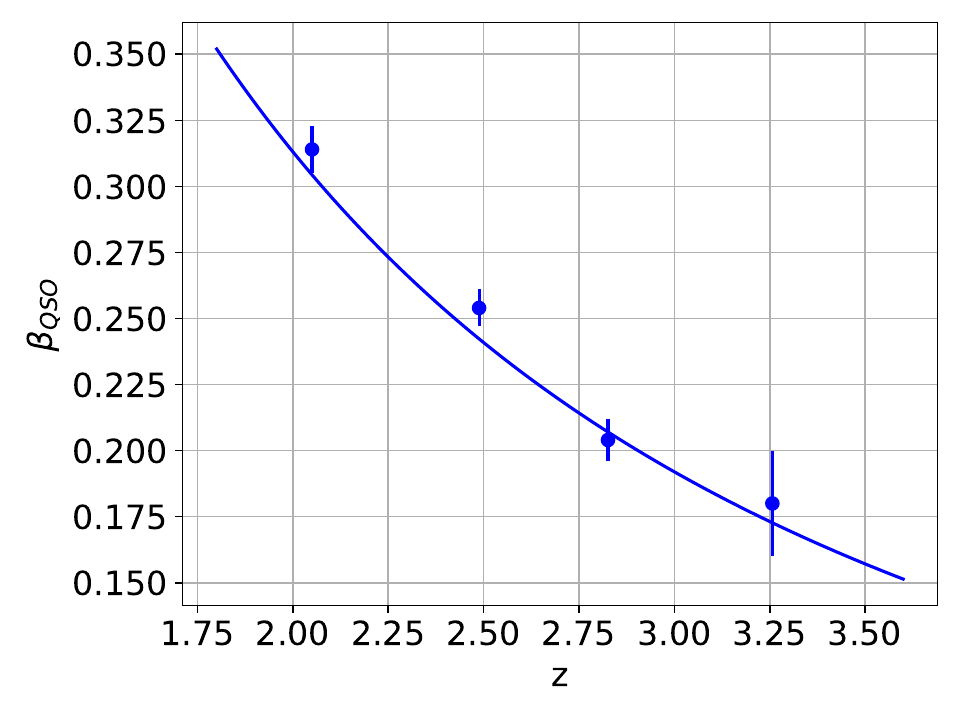} 
\caption{QSO bias (left panel) and QSO RSD parameter (right panel) versus redshift. The continuous lines are the input parameterizations used to build the mocks. 
}
\label{fig:biasQSO}
\end{figure}

The quasar auto-correlation in each redshift bin was fit over the range $20<r<180 h^{-1}$ Mpc, resulting in the bias and RSD parameters displayed in Fig.~\ref{fig:biasQSO}. The $\chi^2$ relative to the input parameterization are 10.8 for the bias and 4.2 for the RSD parameters for four degrees of freedom. The rather large $\chi^2$ for the bias comes from the first two redshift points, which are low by 2.6 and 1.8 $\sigma$, but this corresponds to only 1.7 and 0.9\% of the bias, 
respectively.

\section{Conclusions}
\label{section:conclusions}

We have developed a set of codes to produce \lya\ forest mock data sets using FFT to generate a Gaussian random field with the power spectrum of matter density fluctuations at $z=0$. We use the corresponding linear velocity gradient $\eta$ to implement RSD, which implies producing 6 coordinate fields since $\eta$ is a tensor, in addition to the 3 velocity fields which are needed  for the quasar velocity. This implementation of RSD is done through a modified FGPA, $F= \exp [-a \exp b(\delta +c \eta_\parallel)]$.  A nice feature of such an implementation is that it allows for a prediction (section \ref{sec:prediction}) of the correlation function that is consistent with the measured mock correlation function within eBOSS error bars for $r> 20 h^{-1}$ Mpc. The quasar locations are drawn using interpolation between lognormal density fields at three different redshifts. The production is split into a CPU-expensive pre-production step that draws quasar locations and computes $\delta$ and $\eta_\parallel$ spectra along quasar line-of-sights, and a much faster post-production step, where the FGPA is applied. The code is not parallelized over several CPU nodes, which limits the available memory and then the volume of the data set. With $2.19 h^{-1}$ Mpc voxels, five chunks are needed to cover the NGC and two for the SGC.

The resulting mock data sets generally fit well the desired characteristics with a few issues, typically at $r < 20 h^{-1}$ Mpc. The mean transmitted flux fraction, $\Fbar(z)$ is close to the input parameterization (Fig.~\ref{fig:Fmean}). 
The quasar autocorrelation monopole is 
well-described be linear theory
while the quadrupole is somewhat too large for $r < 20 h^{-1}$ Mpc (Fig~\ref{fig:xiQSO}). The evolution of the quasar bias and the RSD parameter follow the input parameterizations (Fig.~\ref{fig:biasQSO}). The \lya\ autocorrelation in four bins in $z$, times four wedges in $\mu=r_\parallel/r$, is consistent with the prediction (section \ref{sec:prediction}) down to $r=20 h^{-1}$ Mpc and is well fit by the analysis pipeline model down to $r\approx4h^{-1}$Mpc (Fig.~\ref{fig:xi_lya}). 

We aimed at reproducing the bias, $b_{eff}(z)$, and RSD parameter, $\beta(z)$, measured in eBOSS DR16 data and our raw 
and cooked mocks do reproduce these dependencies at the 5\% level
for the auto-correlation (Fig.~\ref{fig:b_lya}). 
For the cross-correlation, we also obtained good agreement
for $b_{eff}(z)$ but found a 10\% disagreement for $\beta(x)$
(Fig.~\ref{fig:xb_lya})

\section*{Acknowledgements}

It is a pleasure to thank Nicolas Busca, Helion du Mas des Bourboux and Anze Slosar for important contributions
during the early stages of this work.  We thank Julian Bautista and the anonymous referee for
comments on the manuscript.

This material is based upon work supported by the U.S. Department of Energy (DOE), Office of Science, Office of High-Energy Physics, under Contract No. DE–AC02–05CH11231, and by the National Energy Research Scientific Computing Center, a DOE Office of Science User Facility under the same contract. Additional support for DESI was provided by the U.S. National Science Foundation (NSF), Division of Astronomical Sciences under Contract No. AST-0950945 to the NSF’s National Optical-Infrared Astronomy Research Laboratory; the Science and Technology Facilities Council of the United Kingdom; the Gordon and Betty Moore Foundation; the Heising-Simons Foundation; the French Alternative Energies and Atomic Energy Commission (CEA); the National Council of Science and Technology of Mexico (CONACYT); the Ministry of Science and Innovation of Spain (MICINN), and by the DESI Member Institutions: \url{https://www.desi.lbl.gov/collaborating-institutions}. Any opinions, findings, and conclusions or recommendations expressed in this material are those of the author(s) and do not necessarily reflect the views of the U. S. National Science Foundation, the U. S. Department of Energy, or any of the listed funding agencies.

The authors are honored to be permitted to conduct scientific research on Iolkam Du’ag (Kitt Peak), a mountain with particular significance to the Tohono O’odham Nation.

\section*{Data Points}

The data points corresponding to each figure in this paper can be accessed in the Zenodo
repository at \url{https://doi.org/10.5281/zenodo.8434293}.

\begin{appendix}

\section{ The z-dependence of quasar correlation function}
\label{app:A}

As discussed in section~\ref{sec:qso}, drawing quasars proportional to the field $\exp (\delta_q)$ results in a correlation function $\xi_0(r)=b_q^2(z_0)G^2(z_0)\xi_m(r,z=0)$.
We aim at producing quasars with a model correlation function that depends on redshift, $\xi_{mod}(r,z)=b_q^2(z)G^2(z)\xi_m(r,z=0)$. 
This dependance can be introduced by drawing quasars with a probability proportional to the field $p=\exp [a(z) \delta_q]$ with 
$a(z) =  \sqrt{\xi_{mod}(r,z) / \xi_0} = b_q(z)(1+z_0) /[b_q(z_0)(1+z)]$, which results in a correlation function $\xi_p(r,z) = \exp [a^2(z) \xi(\delta_q)] -1$. Since $\xi_0=\xi[\exp\delta_q]=\exp[\xi(\delta_q)] -1$ we have
\beq
\xi_p(r,z) = \exp [a^2(z) \ln(1+\xi_0)] -1 = (1+\xi_0)^{a^2} - 1 \approx  a^2 \xi_0 \left(1+\frac{a^2-1}{2} \xi_0 \right) \; .
\eeq
If either $\xi_0 \ll 1 $ or $|a-1|\ll 1$, this is producing quasars with a correlation function $a^2(z)\xi_0=\xi_{mod}(r,z)$, which is what we want. The deviations are, however, significant at low values of $r$ when $z$ is significantly different from $z_0$. For instance for $z_0=2.33$ and $z=3.6$ there is a 18\% deviation at $r=10 h^{-1}$ Mpc.

\begin{figure}[H]
  \centering
 \includegraphics[scale=0.5]{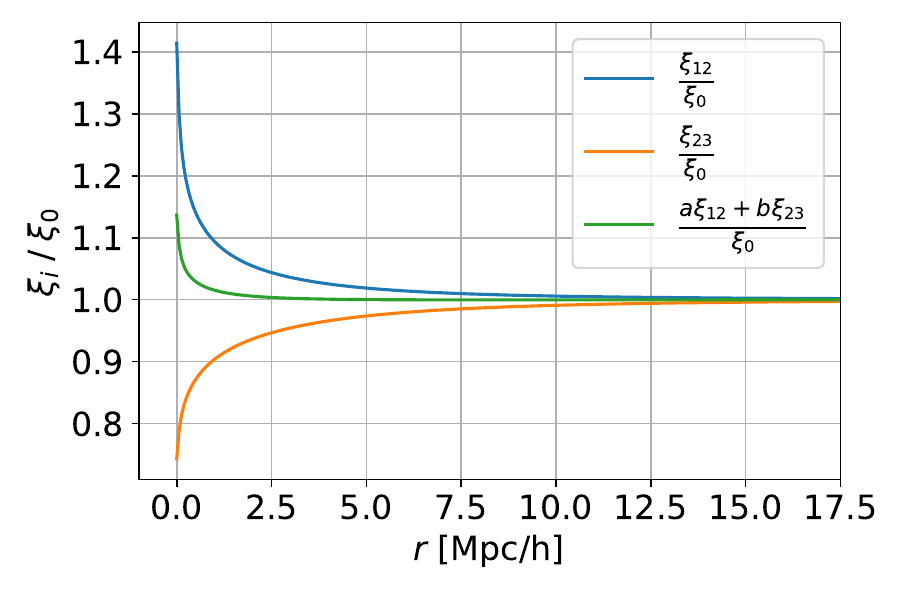} 
 \hspace{1mm} \includegraphics[scale=0.5]{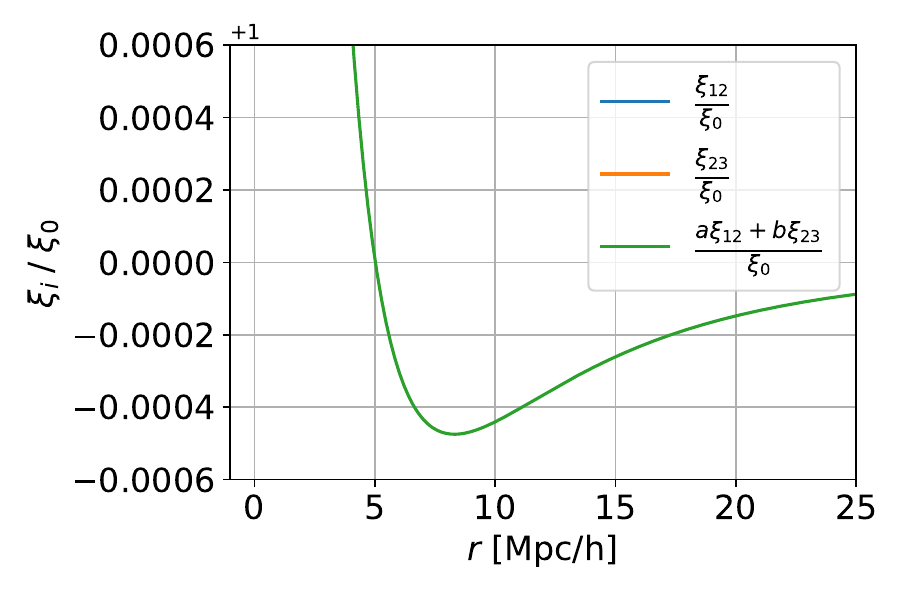} 
  \caption{The ratio $\xi_{12}/\xi_{mod}$  (blue) and $\xi_{23}/\xi_{mod}$ (orange) at $z = 2.38$. For this redshift, $\xi_{12}$ is the linear interpolation of $\xi_1$ and $\xi_2$ and $\xi_{23}$ the linear extrapolation of $\xi_2$ and $\xi_3$. The green curve is obtained as a linear combination of $\xi_{12}$ and $\xi_{23}$, see text for details. The redshift $z = 2.38$ is the redshift for which the green curve deviates the most from $\xi_{mod}$. The right graph is a zoom of the left one. 
  }
  \label{fig:qsolognormal}
\end{figure}

A better result is obtained by producing two quasar-density boxes $\delta_{q1}$ at $z_1$ and $\delta_{q2}$ at $z_2 > z_1$. We define field $p_1=\exp [a_1(z) \delta_{q1}]$ with 
$a_1 = b_q(z)(1+z_1) /[b_q(z_1)(1+z)]$,
which has correlation function $\xi_1$, and define field $p_2$ in a similar way.
At a given $z$ between $z_1$ and $z_2$, we then draw quasars with a probability proportional to the field $p_{12}$ that is the linear interpolation of $p_1$ and $p_2$ as a function of $z$, resulting in a correlation function $\xi_{12}$ that is the linear interpolation of $\xi_1$ and $\xi_2$. This $\xi_{12}$ is slightly lower than $\xi_{mod}(z)$ (orange curve in Fig.~\ref{fig:qsolognormal} left). On the other hand, if we have a third field $p_3$ at $z_3 > z_2$, combining $p_2$ and $p_3$ to extrapolate outside the range $[z_2,z_3]$ down to $z$, results in $\xi_{23}$ that is slightly larger than $\xi_{mod}(z)$ (blue  curve in Fig.~\ref{fig:qsolognormal} left). The shape of these under and over-estimations as a function of $r$ happen to be very close. So finally we produce three quasar-density boxes at $z_1=1.9$, $z_2=2.75$ and $z_3=3.6$.  To produce the probability field at e.g.~$z=2.3$, we combine the field $p_{12}$ from the interpolation between $z_1$ and $z_2$ and the field $p_{23}$ from the extrapolation from $z_2$ and $z_3$, with coefficients that exactly compensate the under and the over-estimates at $r=5 h^{-1}$ Mpc. As a result, we get the model correlation function at better than $5 \times 10^{-4}$ down to $r=5 \, h^{-1}$ Mpc as can be seen in Fig.~\ref{fig:qsolognormal} right.

\section{ Small scale fluctuations}
\label{app:B}

\begin{figure}[H] 
\centering
\includegraphics[scale=0.65]{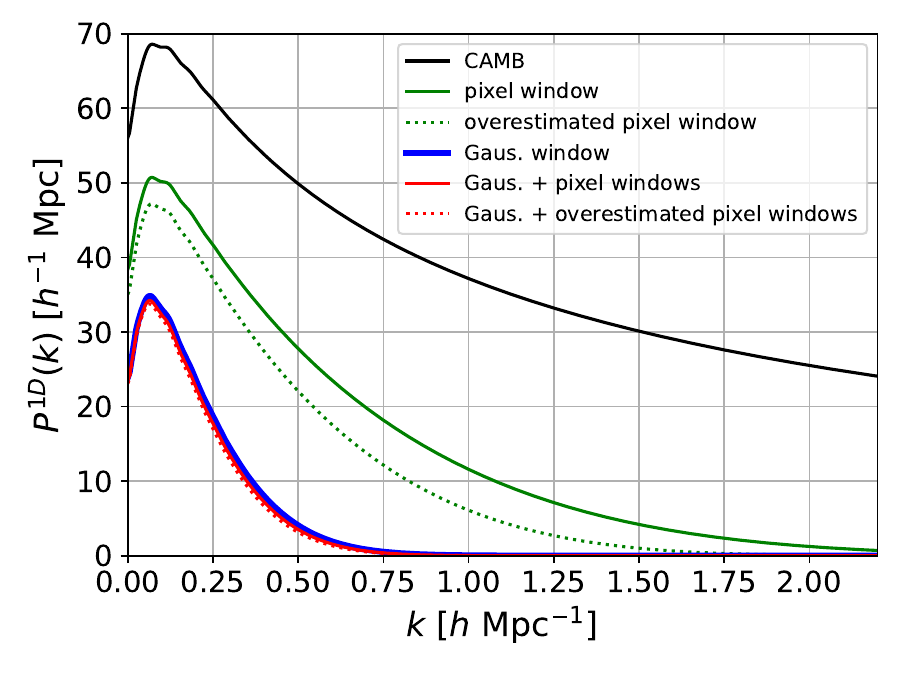} 
\caption{$\Poned$ power spectrum from CAMB (black), same including the voxel size ($a=2.19 h^{-1}$ Mpc) window (green), the Gaussian window (blue) and both of them (red). In the cases where the voxel size window is included (green and red curves), the continuous curves correspond to the approximation $W=\mathrm{sinc}^3(ka/2\sqrt{3})$ (see section \ref{sec:prediction}), while the dotted curve corresponds to the window of a sphere of radius $\sqrt{3}a$, which is underestimating the window function. 
}
\label{fig:P1D}
\end{figure}

The 1d power spectrum is an integral of the 3d power spectrum over $k_\perp$~\citep{KaiserPeacock91},
\beq
\Poned(k_\parallel) = \frac{1}{2\pi}\int_0^\infty P(k_\parallel,k_\perp)k_\perp dk_\perp \; .
\label{eq:P1D}
\eeq
In our mocks $P$ actually includes both the squared voxel window $W^2(k)={\rm sinc}^2(ak_x/2)\times{\rm sinc}^2(ak_y/2){\rm sinc}^2(ak_z/2)$ and the squared Gaussian smoothing window $W^2(k)=\exp(-a^2k^2)$. These window functions cancel 
 high-$k$ contributions to the integral and, since the integral goes to infinity and $P$ is multiplied by a factor $k_\perp$, they strongly reduce $\Poned(k_\parallel)$, as illustrated in Fig.~\ref{fig:P1D}.

It may seem surprising that, as pointed in section 2,  the effect of the window function is hardly visible in a plot of $r^2\xi^{3d}(r)$. The point is that there is a strong effect also on $\xi^{3d}(r)$, in particular $\xi^{3d}(r=0)=\xi^{1d}(r=0)$ is the strongly reduced pixel variance. But since $\xi^{3d}$ is the Fourier transform of $P^{3d}$, when $r$ is large the factor $\exp(-ikr)$ kills the contribution of large $k$ modes for which $W(k)$ is significantly different from unity.
The effect of $W(k)$ is then limited to small $r$ values, which are not visible in a plot of $r^2\xi(r)$ and which are not of interest for BAO analyses. In contrast in Eq \ref{eq:P1D} there is no factor $\exp(ikr)$ and on the contrary there is a factor $k_\perp$, which is increasing the contribution of large $k$ modes.

\end{appendix}

\bibliographystyle{unsrtnat_arxiv}  
\bibliography{saclaymocks}		

\end{document}